\documentclass[aps,pra,twocolumn]{revtex4}
\usepackage{amsfonts}
\usepackage{amssymb}
\usepackage{color}
\usepackage{amsmath}
\usepackage{graphicx}
\usepackage{subfigure}
\usepackage{txfonts}
\usepackage{bm}
\usepackage{ulem}
\usepackage[colorlinks,linkcolor=blue,citecolor=blue,citecolor=blue]{hyperref}

\begin{document}

\title{Quantum non-stationary phenomena of spin systems in collision models}
\author{Yan Li}
\author{Xingli Li}
\author{Jiasen Jin}
\email{jsjin@dlut.edu.cn}
\affiliation{School of Physics, Dalian University of Technology, Dalian 116024, China}

\begin{abstract}
We investigate the non-stationary phenomenon in a tripartite spin-1/2 system in the collision model (CM) framework. After introducing the dissipation through the system-environment collision for both Markovian and non-Markovian cases, we find the emergence of long-time oscillation in the dynamics of the system and the synchronization among subsystems. We connect the CM description and the quantum master equation in the continuous time limit and explain the existence of the stable oscillation by means of Liouvillian spectrum analysis. We investigate the thermodynamics of persistent oscillations in our CM in both Markovian and non-Markovian regimes. In addition, we find that the imperfection of collective dissipation can be compensated by the randomness of the interaction sequence in our CM.
\end{abstract}
\date{\today}
\maketitle

\section{Introduction}
The quantum system that is subjected to coupling to an uncontrollable environment is referred to as an open quantum system. The
inevitable interaction with the environment always leads to the dissipative effect of the open quantum system and the non-unitary feature of its time-evolution \cite{Breuer2002,Rivas2012,Breuer2016}.  In the presence or absence of memory effects, the dynamics of open quantum systems can be classified into the memoryless Markovian process and the non-Markovian master process which allows the backflow of information during the time-evolution. In dissipative open quantum systems, in comparison with their closed counterparts, plenty of intriguing phenomena emerge especially in the quantum many-body systems, such as the steady-state phase transition \cite{Banchi2014,Maile2018,Hwang2018,Prasad2022}, information spreading \cite{Zhangyongliang2019,Styliaris2021,Zanardi2021},  quantum many-body delocalization \cite{Potter2015,Hyatt2017,Rubio2019,Xian2021}, etc. Generally, the state of an open quantum system asymptotically evolves to one or more time-independent steady states in the long-time limit. However, there are exceptional cases in which the system evolves to non-stationary states. The time-dependent long-time behavior of the system intimately related to the quantum time crystals \cite{Wilczek2012,Kozin2019,Georg2021}, quantum chaos \cite{Zhuang2013,Bruzda2010}, and quantum synchronization \cite{Laskar2020PRL,Jaseem2020PRR,Solanki2022PRAL,Hajdusek2022PRL,Krithika2022PRA}.

As the extension of classical synchronization in the quantum regime, quantum synchronization has attracted much attention in the last few years. The quantum synchronization between self-sustained oscillations can be established through either the external driving or the internal coupling \cite{ZhirovPRL2008,ZhirovPRB2009,GalvePRL2010,WalterPRL2014,AmeriPRA2015}. The former is known as the forced synchronization or entrainment \cite{Karpat2019PRA} while the latter is called spontaneous synchronization which stems from the quantum correlations in the systems. Meanwhile, the quantum correlations in the systems is not limited to direct interactions between subsystems, but also may emerge indirectly through the interactions of subsystems together with their surrounding environment. This quantum correlations can eventually lead to non-stationary phenomena depicted by the quantum synchronization measure. In recent years, quantum synchronization in open quantum systems has already been explored at length in a variety of quantum systems, such as optomechanical arrays \cite{Cabot2017NJP,Ludwig2013PRL,RodriguesNC2021}, van der Pol (VdP) oscillators \cite{Walter2014PRL,Tilley2018NJP}, atomic ensembles \cite{Xu2014PRL}, and superconducting circuit systems \cite{Vinokur2008Nature,Quijan2013PRL,Fistul2017SR}.

In general, theoretic descriptions for the dynamics of open quantum systems as well as the non-stationary behavior in the long-time limit are carried out by calculating the expectation values of the local observables and correlations over time through the dynamical equations of the system such as the quantum master equation and quantum Langevin equation. In recent years, investigations on the open quantum system within the framework of collision model (CM) are reported \cite{Rau1963,Ziman2002,Scarani2002,Ziman2005,Ciccarello2013PRAR}. The CM approach can provide intuitive pictures of the interactions between the system and its environment as well as the strategies of the information flows in the time-evolution of the state of the system. Recently, it is shown that the CM can efficiently reproduce the dissipative collective phenomena of  multipartite open quantum systems \cite{cattaneo2021prl,cattaneo2022arXiv}. In the CM framework, the coupling of the system and environment is simulated by repeated collisions between system and a set of environmental particles. Since the flexibility and scalability in designing the collision details, the CM becomes a powerful tool for investigating non-Markovian dynamics \cite{Ciccarello2013,McCloskey2014,Jinjiasen2015,Lorenzo2017,Jinjiasen2018,Camasca2021,Filippov2022,Francesco2022,cattaneo2022osid}, quantum information scrambling \cite{Liyan2020,Liyan2022}, quantum steering \cite{Beyer2018}, quantum friction \cite{Grimmer2019}, multipartite entanglement generation \cite{celse2019}, and quantum synchronization \cite{Karpat2019PRA, Karpat2021} in complicated open quantum systems. Recently, the experimental realization of an all-optical collision model has already been demonstrated \cite{Cuevas2019}.

In this work, we utilize the CM to investigate the long-time behavior of a composite spin system consisting of three spin-1/2 subsystems subjected to thermal environment. By varying the strength of the interactions within the environment blocks, we can obtain the Markovian dynamics and non-Markovian dynamics, respectively. We find that, without additional external driving, the system is able to reach a stable non-stationary steady state only through its internal interactions and dissipative processes. At this point synchronization phenomena are also constructed between subsystems which reminisce the results reported by Karpat, {\it et al.} in Ref. \cite{Karpat2020PRA}. Furthermore, after taking the continuous time limit for the CM, we establish the connection between the CM and the Markovian master equation in Lindblad form. We interpret the appearance of the underlying long-time oscillations of the subsystems by analyzing the Liouvillian spectrum of the associated quantum master equation. In addition, we investigate the thermodynamical properties and correlations between the system and the environment when the dynamics of the subsystems are synchronized.

The paper is organized as follows. We first explain the idea of the our specific model in the CM framework and its utility in simulating the dynamics of the composite systems consisting of three spins in Sec.\ref{model}. In Sec.\ref{Markovian case} we then focus on the Markovian dynamics to show the temporal expectation values of the local observables. In particular for the case that the subsystems enter into the long-time oscillations we discuss the quantum synchronization among the subsystems. We explore the connection between the CM and Lindblad master equation in describing the underlying system in the continuous time limit. We present our understanding on the appearance of oscillations in the dynamics from the Liouvillian spectrum. The thermodynamical properties and the effects of imperfect collective dissipation on the dynamics. In Sec. \ref{non-Markovian case}, we discuss the time-evolution of the systems in the non-Markovian case. Finally, we summarize in Sec. \ref{Summary}.

\section{The framework of Collision model}
\label{model}

\begin{figure}[!htpb]
  \includegraphics[width=1\linewidth]{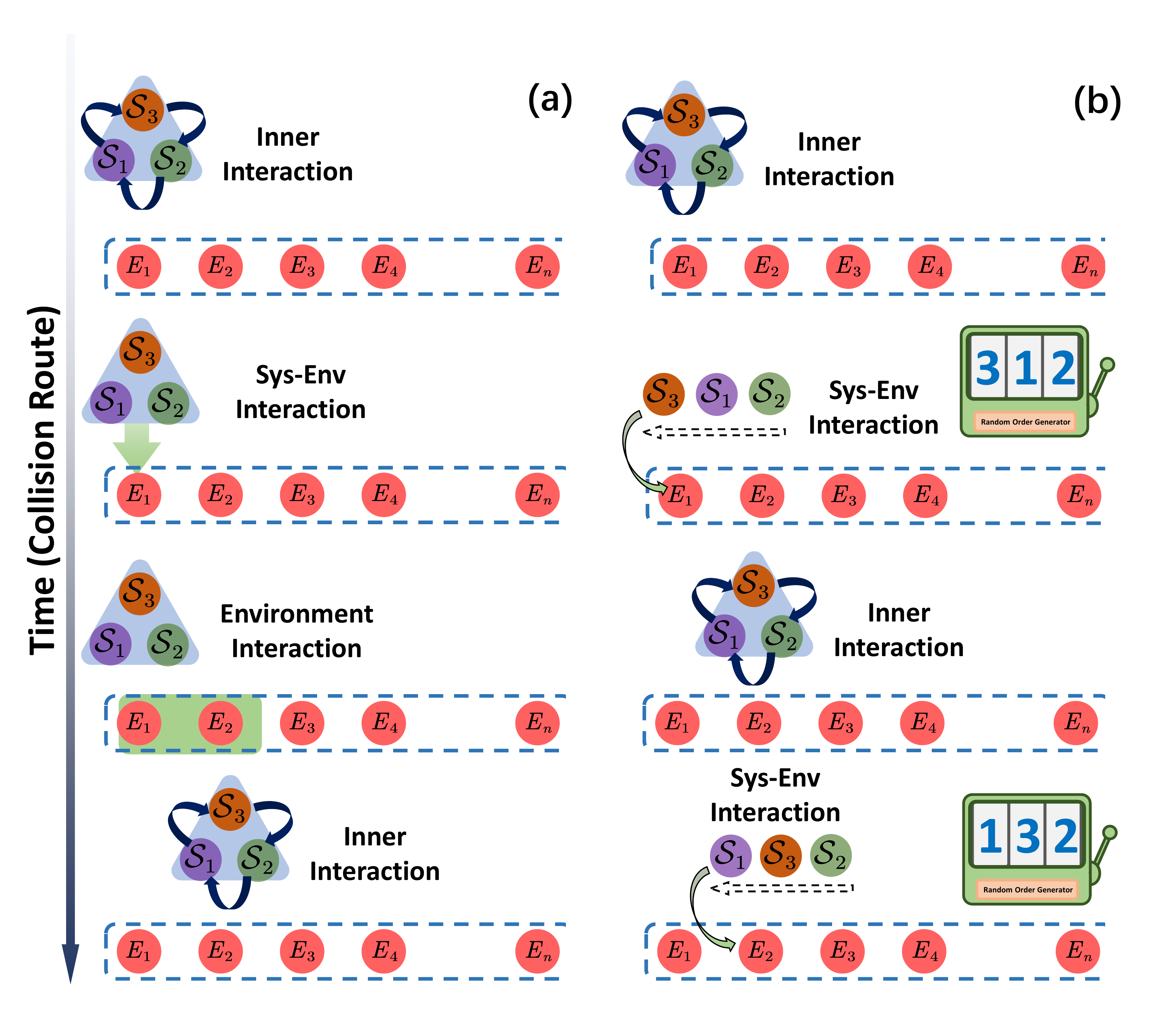}
  \caption{\label{FIG:CMSchematic} Schematic of routes of collision models: (a) Collective dissipation: At the end of the interaction within the system, the tripartite system collides with the environment spin $E_{n}$ simultaneously. (b) Random local dissipation. In this case, the subsystems collide with the environment blocks one by one and the interaction order is randomly determined each time.}
\end{figure}

In the generic CM framework, the entire installation consists of two parts: the system $\mathcal{S}$ and its environment $E$. Usually, the environment is represented by a set of identical particles denoted by $\{E_{1},E_{2},\cdots, E_{j},\cdots\}$ which are initialized in the same state. The interaction between the system and the environment is simulated by the successive collisions between system particles and environmental particles in a stroboscopic manner. In our CM, both the system and environment are consisted of a set of spin-$1/2$ particles. In particular, the system of interest is tripartite with interacting subsystems $\mathcal{S}_1$, $\mathcal{S}_2$ and $\mathcal{S}_3$.
The inner interactions between the subsystems are given by (set $\hbar=1$ hereinafter),
\begin{equation}
H_{\mathcal{S}} = \sum_{\alpha} \sum_{m=1}^3{ J_{\alpha}\sigma_{m}^{\alpha} \sigma_{m+1}^{\alpha}},
\label{Hss}
\end{equation}
where $\sigma_{m}^{\alpha}$ ($\alpha = x,y$ and $z$) are the Pauli matrices of the $m$-th subsystem and $J_{\alpha}$ is the coupling strength between system spins. Notice that the periodic boundary condition is imposed in Eq. (\ref{Hss}).

We consider the case that the environmental spins are uncorrelated and are prepared in the identical thermal state $\eta^{j}_{\text{th}}=e^{-\beta H_j}/\text{tr}(e^{-\beta H_j})$, $\forall j$. The free Hamiltonian of each environmental particle is $H_j = \omega \sigma_j^{z}/2$ where $\omega$ describes the self-energy and $\beta=1/k_BT$ is the inverse of temperature. Although the environment spins are initialized in the uncorrelated state, interactions between environment particles are allowed to take place. As will be seen in the next sections, it is the intra-environment interactions that generate the memory effect of the stroboscopic evolution of the system.

Now let us illustrate the setup of our CM which is schematically shown in Fig. \ref{FIG:CMSchematic}(a). The CM works through the following steps:

\noindent {\it \textbf{Step 1.}} The CM starts with the collisions among the subsystems $\mathcal{S}_1$, $\mathcal{S}_2$ and $\mathcal{S}_3$
according to Eq. (\ref{Hss}). The time-evolution of the state of the system is then described by the unitary operator
$U_{\mathcal{S}}$ = $\exp (-i H_{\mathcal{S}} \tau_{\mathcal{S}})$ and $\tau_{\mathcal{S}}$ is the interaction time.

\noindent {\it \textbf{Step 2.}} Collisions take place between the system $\mathcal{S}$ and environmental spin $E_{n}$.
In this step, the $\mathcal{S}$-$E_{n}$ collision simulates the interaction between the system and environment which is specified by the flip-flop Hamiltonian in our CM as follows,
\begin{equation}
H_{I} = g\sum_{m} \sigma_{m}^{-}\sigma_{E_{n}}^{+} + \sigma_{m}^{+}\sigma_{E_{n}}^{-} ,
\label{Hse}
\end{equation}
where $g$ denotes the coupling strength of system-environment interaction. The interaction shown in Eq. (\ref{Hse}) describes the collective flip-flop process. This interaction is known as a good description for the collective spontaneous emission (Dicke superradiance) \cite{AndreevSPU1980,Levitt2012,MassonNC2022}. From an experimental point of view, this collective decay can be realized on the trapped-ion platform and mediated by an auxiliary ion \cite{Schneider2002PRA,LeePRA2014}. The corresponding time-evolution operator is $U_{I} = e^{-iH_{I}\tau_{I}}$ where $\tau_{I}$ denotes the interaction time.

\noindent {\it \textbf{Step 3.}} The collision takes place between the environment particles $E_{n}$ and $E_{n+1}$ which may induce non-Markovian dynamics.
The interaction Hamiltonian for the $E_{n}$-$E_{n+1}$ interaction reads
\begin{equation}
H_{E} = g_{E}\sum_{\alpha}\sigma_{E_n}^{\alpha} \sigma_{E_{n+1}}^{\alpha},
\label{Hee}
\end{equation}
where $g_E$ is the coupling strength between environment spins. Thus the unitary time-evolution operator can be expressed as $U_{E}$ = $e^{-iH_{E}\tau_{E}}$ with $\tau_E$ being the corresponding interaction time. Eq. (\ref{Hee}) can be reexpressed in the form of a partial SWAP operation,
\begin{equation}
U_{E} = \cos{\theta} \mathbb{I}_{4} + i\sin{\theta}U_{\text{SWAP}},
\label{Eq:UEE}
\end{equation}
where $\mathbb{I}_{4}$ is the $4\times 4$ identity operator and the SWAP operation is given by $U_{\text{SWAP}} = \left| 00 \rangle \langle 00 \right|$ + $\left| 01 \rangle \langle 10 \right|$ + $\left| 10 \rangle \langle 01 \right|$ + $\left| 11 \rangle \langle 11 \right|$.
The parameter $\theta=2g_{E}\tau_{E}$ controls the strength of the SWAP operation with $\theta\in[0,\pi/2]$. The non-Markovianity of the system dynamics can be switched on by tuning the parameter $\theta$.

After the $E_{n}$-$E_{n+1}$ collision, the spin $E_{n+1}$ carries part of the information that flows into the environment in step 2. The system together with the $(n+1)$-th environment particle enters into the next loop and the spin $E_n$ is discarded as shown in Fig.\ref{FIG:CMSchematic}(a). Therefore, the state of the system of interest after the $n$-th loop ($n\ge2$) is transformed into
\begin{equation}
\rho^{n}_{\mathcal{S}}\mapsto \rho^{n+1}_{\mathcal{S}} = \text{tr}_{E_n,E_{n+1}}\left[U_EU_IU_{\mathcal{S}}\left(\rho^{n}_{\mathcal{S}E_n}\otimes\eta^{n+1}_{\text{th}}\right)U_{\mathcal{S}}^\dagger U_I^\dagger U_E^\dagger\right],
\label{rhotot}
\end{equation}
where $\rho^{n}_{\mathcal{S}E_n}$ is the joint state of the system and the $n$-th environment unit after collision. The system is initialized in the state $\rho^{\text{ini}}_{\mathcal{S}}=|\psi\rangle_{\mathcal{S}}\langle \psi|$ with the separable state $|\psi\rangle_{\mathcal{S}}=|\psi\rangle_{\mathcal{S}_{1}}\otimes|\psi\rangle_{\mathcal{S}_{2}}\otimes|\psi\rangle_{\mathcal{S}_{3}}$. More precisely, since the Hamiltonian has the symmetry along the spin $z$-axis we choose the initial state of each subsystem as the $120^{\circ}$-state on the equatorial plane of Bloch sphere, i.e., $\psi_{\mathcal{S}_m}=\left(|\uparrow\rangle+e^{im\frac{2}{3}\pi}|\downarrow\rangle\right)/\sqrt{2}$, with $m=1,2$, and $3$. The chosen $120^{\circ}$-state will facilitate the discussion on the evolution of the phase differences among the state of subsystems. For simplicity, we set the interaction time for the each collision as $\tau=\tau_{\mathcal{S}}=\tau_{I}=\tau_{E}$ in the rest of the analysis.

\section{Markovian case}
\label{Markovian case}
In this section, we focus on the case that the intra-environment collision is absent in the CM, i.e. $U_{E}$ reduces to an identity matrix implying that the dynamics of the system of interest is Markovian. We choose the temporal expectation value of local observable $\langle\sigma^{x}_{m}\rangle=\text{tr}(\sigma^{x}_{m}\rho_{\mathcal{S}})$ to monitor the dynamics of the system. We first focus on the effect of the environmental temperature $\beta$ on the time-evolution of the state of the system. Note that $\beta$ is proportional to $1/T$.

\begin{figure}[!htpb]
  \includegraphics[width=1.05\linewidth]{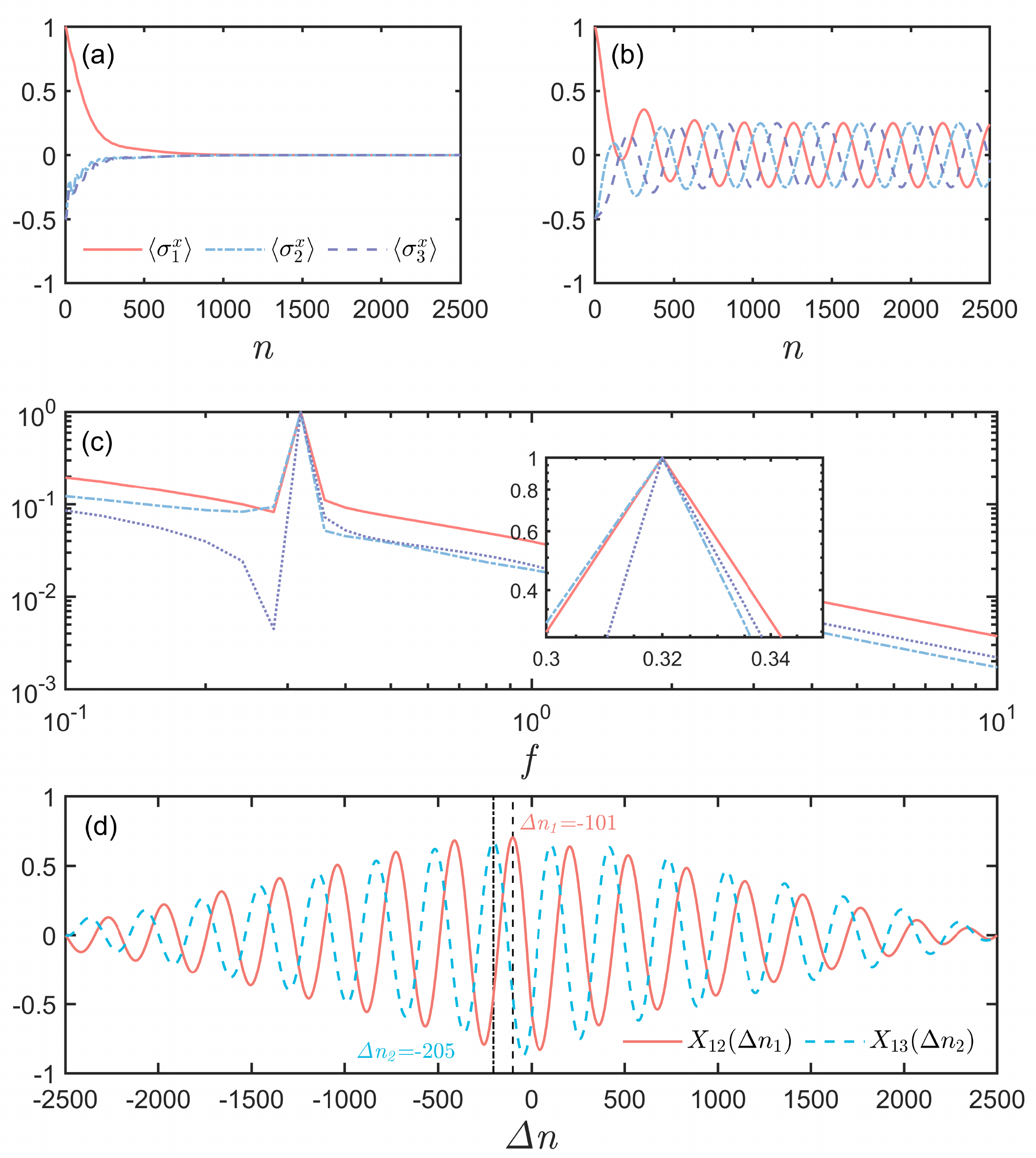}
  \includegraphics[width=1.05\linewidth]{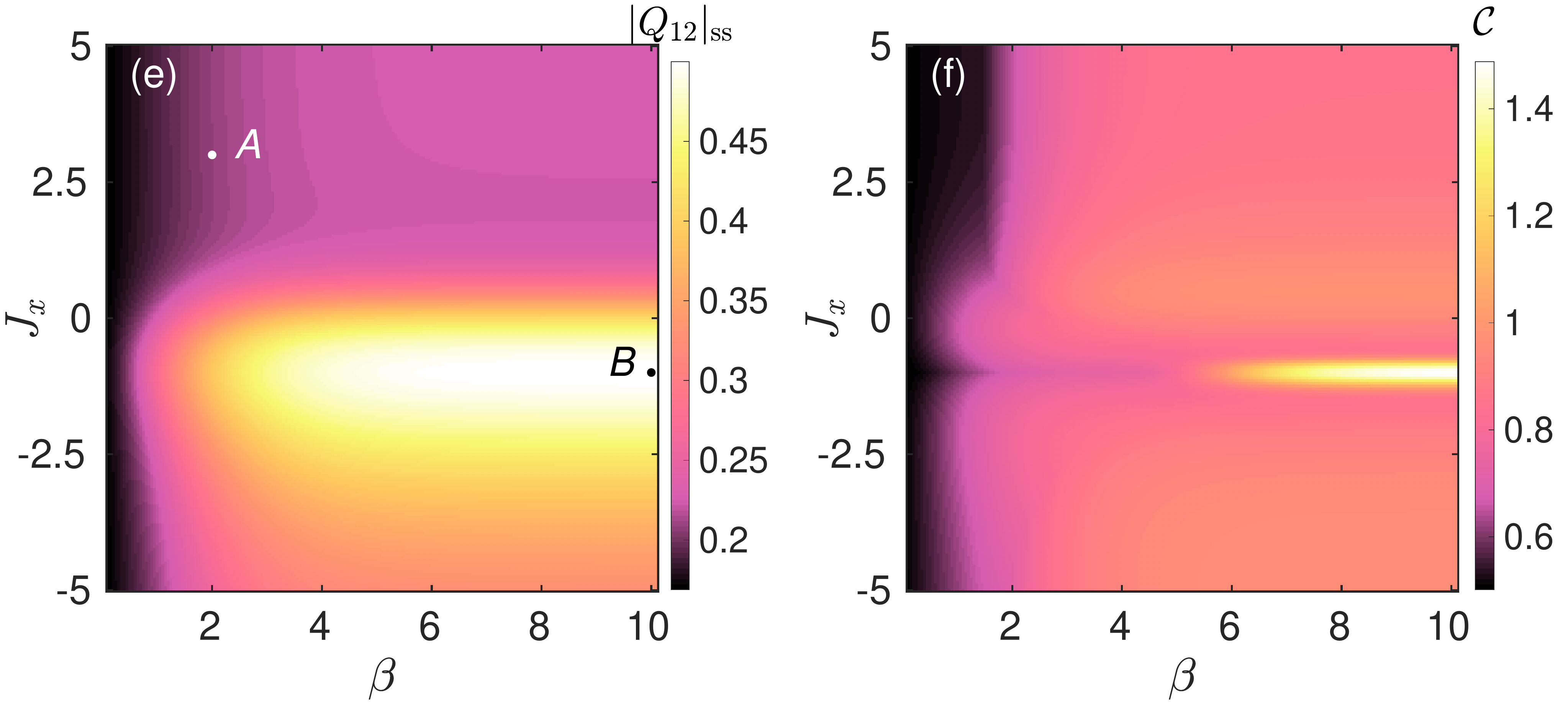}
  \caption{\label{FIG:SyncVSCoh} The stroboscopic time-evolution of the local observables $\langle\sigma^{x}_{m}\rangle$ ($m=1,2,3$) with  $J_{x}=3, \beta=2$ (a) and $J_{x}=-1, \beta=10$ (b). (c) The Fast Fourier Transform for the $\langle\sigma^{x}_{m}\rangle$s shown in panel (b), the zoom-in hints those three local observables sharing an identical dominant frequency $f^{d}\approx0.32$. (d) The cross-correlations $X_{1k},(k=2,3)$ of the $\langle\sigma^{x}_{m}\rangle$s was shown in panel (b). (e) The absolute value of the synchronization measure $Q_{12}(n)$ (computed up to $n=20000$ collisions) as a function of the environment temperature $\beta=1/k_BT$ and the coupling strength $J_{x}$. The black dots $A$ and $B$ mark the parameters ($J_{x}=3, \beta=2$) and ($J_{x}=-1, \beta=10$), respectively. (f) The quantum coherence $\mathcal{C}$ in the system, as measured by the $l_{1}$ norm of coherence, and the parameters are chosen the same as in panel (e).  Other parameters in all the pannels are chosen as $J_{y}=-1,J_{z}=1,g=10$ and $\tau=0.01$. }
\end{figure}

In Figs.\ref{FIG:SyncVSCoh}(a) and (b), we show the time-dependence of the local observables $\langle\sigma^{x}_{m}\rangle$ for different $\beta$. One can see that the magnetizations of all the spins trivially approach to zero after a sufficient large number of collisions
for $J_{x}=3$ and $\beta=2$. As the environmental temperature goes down, the asymptotically values of the magnetiaztions of all the spins become unstable and eventually the time-evolution of the magnetizations enter into the oscillating trajectories. The time-dependent oscillations of $\langle\sigma^{x}_{m}(n)\rangle$ for $\beta=10$ can be observed in Fig.\ref{FIG:SyncVSCoh}(b). Moreover, as shown in Fig.\ref{FIG:SyncVSCoh}(c), the Fast Fourier Transform analysis on the numerics of $\langle\sigma^{x}_{m}(n)\rangle$ reveals that the oscillations share the same dominant frequency. A zoom-in at the vicinity of the peak in the frequency domain shows that the dominant frequency is $f^{d}\approx0.32$.

The phase-locking feature of the oscillations can be verified by checking the cross-correlation between the time-evolutions of $\langle\sigma^x_m(n)\rangle$. Generally, the cross-correlation quantifies the degree of the association between these two time-dependent functions $f(n)$ and $g(n)$ and is defined as their convolution-like function.  Here, we analogously express the cross-correlation for the discrete evolutions of $\langle\sigma^x_m(n)\rangle$ as
\begin{equation}
X_{jk}(\Delta n) = \sum_{n=1}^N{\langle\sigma^x_j(n)\rangle\langle\sigma^x_k(n-\Delta n)\rangle},
\label{Xcorr}
\end{equation}
where $\Delta n$ is the amount of translation and $N$ is the total number of collisions. The cross-correlation function is actually an inner product of two vectors representing the projection of one onto another in the linear space. Therefore the cross-correlation can faithfully capture the similarity of two vectors under different amounts of translation within a period. The maximum of $X_{jk}(\Delta n)$ in Eq. (\ref{Xcorr}) reveals the optimal translation that shifts the trajectory of $\langle\sigma^x_j\rangle$ closest to $\langle\sigma^x_k\rangle$ and thus in turn reflects the phase difference $\Delta\phi_{\mathcal{S}_j\mathcal{S}_k}$. In Fig. \ref{FIG:SyncVSCoh}(d), we show the cross-correlation $X_{jk}(\Delta n)$ produced by the data in Fig. \ref{FIG:SyncVSCoh}(b), one can see that the maxima of the cross-correlation $X_{12}(\Delta n_1)$ and $X_{13}(\Delta n_2)$ appear at $\Delta n_{1}=-101$ and $\Delta n_{2}=-205$, respectively. Recall the previously calculated dominant frequency $f^d\approx 0.32$, the phase differences can be obtained as $\Delta\phi_{\mathcal{S}_{1}\mathcal{S}_{2}}=2\pi f^{d}/\Delta n_{1}\tau\approx-2\pi/3$, $\Delta\phi_{\mathcal{S}_{1}\mathcal{S}_{3}}=2\pi f^{d}/\Delta n_{2}\tau\approx-4\pi/3$. The initial phase differences between the sublattices are conserved after the dissipative evolution. Note that we have rescaled the number of collisions $n$ into time $n\tau$ with time interval $\tau=0.01$ .

The phase-locking oscillation in the long-time dynamics of the system indicates the emergence of the spontaneous synchronization among the subsystems. In the following we will employ the nonlocal synchronization measure proposed by Es'haqi-Sani {\it et al.} in Ref. \cite{NajmehPRR2020} to quantify the synchronization features in our CM. The mentioned measure is a temporal complex-valued correlator and defined as follows,
\begin{equation}
Q_{jk}(n)=\frac{\langle\sigma^{+}_{j}\sigma^{-}_{k}\rangle_{n}}{\sqrt{\langle\sigma^{+}_{j}\sigma^{-}_{j}\rangle_{n}\langle\sigma^{+}_{k}\sigma^{-}_{k}\rangle_{n}}},
\end{equation}
where $\langle O \rangle_{n}=\text{Tr}[O\rho^{n}_{\mathcal{S}}]$ is the expectation value of the observable $O$ after the $n$-th collision. The modulus of $Q_{jk}(n)$ characterizes the degree of non-local correlation, for instance, two subsystems are completely correlated when $|Q_{jk}(t)|\to1$. Combining the dynamics of local observable of the system and the behavior of this non-local correlation, we are able to investigate the synchronization properties. When the dynamics of the subsystems are oscillating and the non-local correlation function tends to a stable value, the subsystem can be considered to be synchronized. In Fig. \ref{FIG:SyncVSCoh}(e), we show the modulus for the steady-state $|Q_{12}|_{\text{ss}}$ in the $\beta$-$J_{x}$ plane.
One can see that the lower environmental temperature ($\beta\rightarrow\infty$) could facilitate the synchronization of among the subsystems. Moreover, for $\beta\gtrsim 4$ the long-time oscillation of the subsystems becomes almost fully synchronized at an optimal interaction strength $J_x=J_y$. Namely, the anisotropy of the system Hamiltonian tends to drive the dynamics of the subsystems far way from the synchronization.

On the other hand, the coherence property is also related to synchronization \cite{Galve2017} which has already been considered as a synchronization measure \cite{Jaseem2020PRE,Karpat2020PRA}. In Fig. \ref{FIG:SyncVSCoh}(f) we present the $l_1$ norm of coherence of the state of the system in the $\beta-J_{x}$ plane. The $l_1$ norm of coherence is defined as follows,
\begin{equation}
\mathcal{C}=\sum_{p\neq q}|\langle p| \rho_{S}|q\rangle|,
\end{equation}
where $\{|p\rangle\}$ are the computational basis of $\rho_{\mathcal{S}}$ \cite{Baumgratz2014PRL}. Indeed, we find that there are many similar behaviors in Figs. \ref{FIG:SyncVSCoh}(e) and (f). For lower environmental temperature or an anisotropic Hamiltonian, the coherence of the system are poorly present in the system which implies that the synchronization is more likely to be established when the state of system $\rho_{\mathcal{S}}$ contains more coherence.

\begin{figure*}[!htpb]
  \includegraphics[width=1\linewidth]{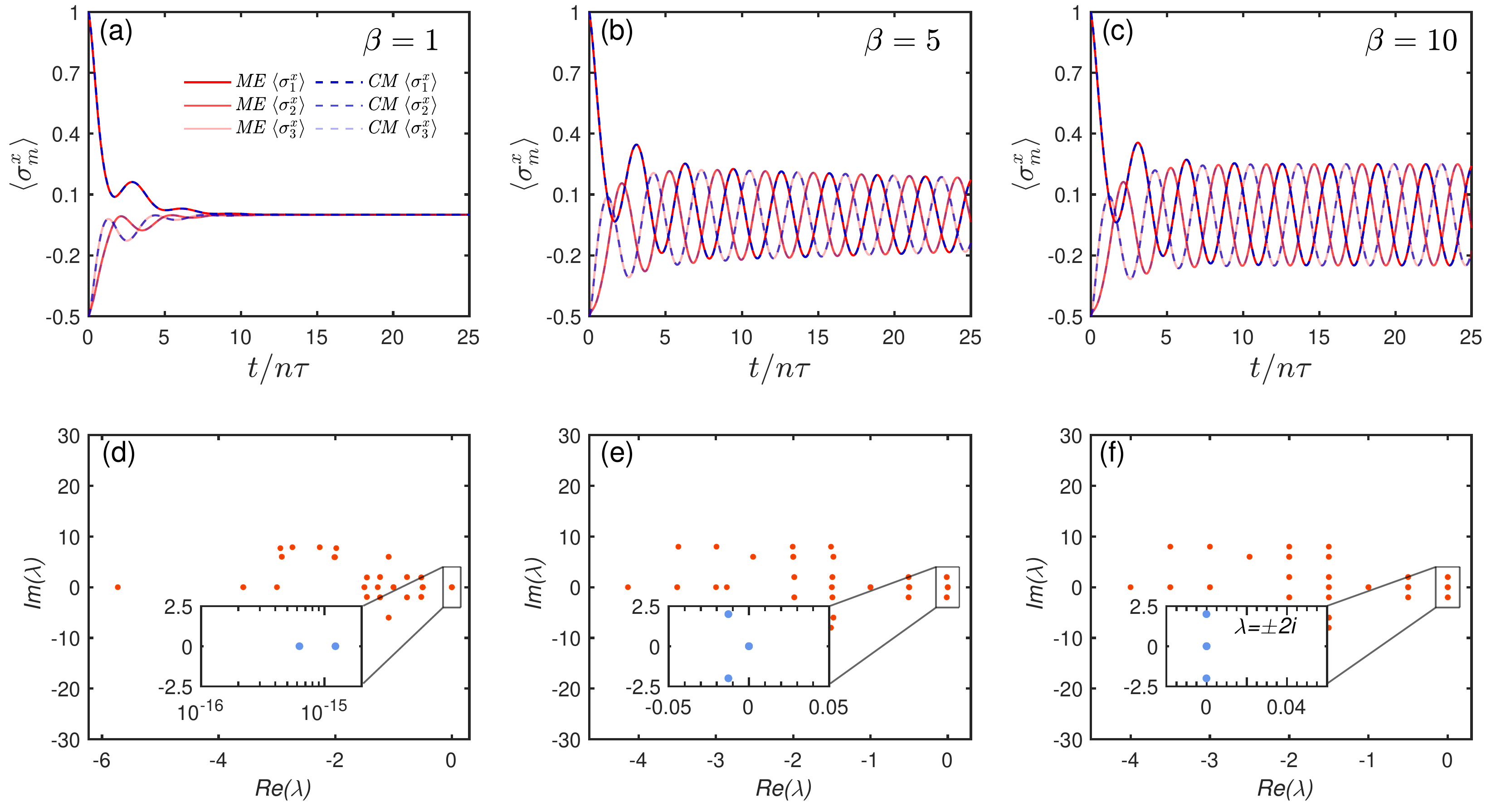}
  \caption{\label{FIG:MEvsCM} Top panels: the time-evolution of expectation values $\langle\sigma^{x}_{m}\rangle$ $(m = 1, 2, 3)$ of the system $\mathcal{S}$ (different subsystems are distinguished by the transparency of the line) in the description of master equation (solid lines) and collision model (dashed lines). Bottom panels: the Liouvillian spectra in the complex planes. The environmental temperatures are $\beta=1$ [(a) and (d)], $5$ [(b) and (e)] and $10$ [(c) and (f)]. All the zoom-in show the details of the eigenvalues around the zero real part. Other parameters are chose as $J_{x}=J_{y}=-1,J_{z}=1,\gamma=1,g=10$ and $\tau=0.01$ for all the panels. }
\end{figure*}

\subsection{Continuous time limit}
\label{continuoustimelimit}
In the preceding, we have found that the emergence of the synchronization depends on both the properties of the system Hamiltonian and the environment temperature. In order to figure out the underlying physics, we are going to analyze the dynamics of the system by the taking the continuous time limit for the Markovian CM. We start with a discussion on a single loop, say the $n$-th, in a more general CM with the system-environment interaction Hamiltonian in the form of $V=g\sum_{k}\mathcal{A}_{k}\otimes E_{k}$ where $\mathcal{A}_{k}$ and $E_{k}$  are the operators of the system and environment spins. Notice that since we only concern with the final state of the system after the $n$-th loop in the Markovian CM, the collision between $E_n$ and $E_{n+1}$ is not considered for the moment. The map for the system state is thus given by,
\begin{equation}
\rho_{\mathcal{S}}^{n}\mapsto \rho_{\mathcal{S}}^{n+1} = \text{tr}_{E_{n+1}}\left[U_{I}U_{\mathcal{S}} (\rho^{n}_{\mathcal{S}} \otimes \eta^{n+1}_{\text{th}})U_{\mathcal{S}}^{\dagger}U_{I}^{\dagger}\right]=\Lambda[\rho^{n}_{\mathcal{S}}],
\label{Eq:CMofStateMap}
\end{equation}
where $\Lambda[\cdot]$ is a completely positive trace-preserving (CPTP) map acting on the system density matrix $\rho_{\mathcal{S}}$ and $U_{I}=e^{-iV\tau}$ is a unitary operator describing the collision between the system and environment. In the continuous time limit ($\tau\rightarrow 0$), we can expand the unitary operators as follows,
\begin{equation}
U_{I}=\mathbb{I}-i\tau V-\tau^2 V^2/2+o(\tau^n),
\label{Use_exp}
\end{equation}
and
\begin{equation}
U_{\mathcal{S}}=\mathbb{I}-i\tau H_{\mathcal{S}}+o(\tau^n).
\label{Uss_exp}
\end{equation}
Therefore, the variation of $\rho_{\mathcal{S}}$ between two consecutive loops can be obtained as follows,
\begin{equation}
\delta \rho_{\mathcal{S}}=\rho_{\mathcal{S}}^{n+1} -\rho_{\mathcal{S}}^{n} = (\Lambda-\mathbb{I})[\rho^{n}_{\mathcal{S}}],
\label{Eq:dynmapping}
\end{equation}
By substituting Eqs. (\ref{Use_exp}) and (\ref{Uss_exp}) into Eq. (\ref{Eq:dynmapping}), we have
\begin{equation}
\begin{aligned}
\delta \rho_{\mathcal{S}}\approx& -i\tau[H_\mathcal{SS},\rho^{n}_\mathcal{S}]+\frac{\tau^2}{2}\text{tr}_{E}(2V\rho^{n}_\mathcal{S}\otimes\eta^{n+1}_{\text{th}}V-\{V^2,\rho^{n}_\mathcal{S}\otimes\eta^{n+1}_{\text{th}}\})\\
&-i\tau\text{tr}_{E}([V,\rho^{n}_\mathcal{S}\otimes\eta^{n+1}_{\text{th}}]) -\tau^{2}\text{tr}_{E}(\{\rho^{n}_\mathcal{S}\otimes\eta^{n+1}_{\text{th}},H_{\mathcal{S}}V\})\\
&-\tau^{2}\text{tr}_{E}(V\rho^{n}_\mathcal{S}\otimes\eta^{n+1}_{\text{th}}H_{\mathcal{S}}-H_{\mathcal{S}}\rho^{n}_\mathcal{S}\otimes\eta^{n+1}_{\text{th}}V).
\label{eq:CMtoME}
\end{aligned}
\end{equation}
Under the assumptions $\text{tr}_{E}[\eta^{n+1}_{\text{th}} V]=0$ and $\text{tr}_{E}[\eta^{n+1}_{\text{th}} V^2]\neq0$, which is trivially satisfied in the cases that the initial environment states have zero first-order moment, we divide both sides of Eq. (\ref{eq:CMtoME}) by the collision time $\tau$ and may obtain the following quantum master equation in Lindblad form,
\begin{equation}
\frac{d}{dt}\rho_\mathcal{S}(t) = -i[H_\mathcal{S},\rho_\mathcal{S}]+\frac{g^2\tau}{2}\sum_{k,l}\Gamma_{k,l}(2\mathcal{A}_{k}\rho_\mathcal{S}\mathcal{A}_{l}-\{\mathcal{A}_{l}\mathcal{A}_{k},\rho_\mathcal{S}\}),
\label{Eq:CTLME}
\end{equation}
where $\Gamma_{k,l}= \langle E_{l}E_{k}\rangle_{\eta^{n+1}_{\text{th}}}$ is the environment correlation function and the dimensionless parameter $g^2 \tau$ is the effective decay rate. In our CM, the jump operators in Eq. (\ref{Eq:CTLME}) are determine to be the collective lowering and raising operators $A^{\pm}=\sum_m{\sigma^{\pm}_{m}}$, we end up with the master equation in the Lindblad form,
\begin{equation}
\begin{aligned}
\frac{d}{dt}\rho_\mathcal{S}(t) = & -i[H_{\mathcal{S}},\rho_\mathcal{S}(t)]\\
&+ \frac{\gamma(1-\xi)}{4}(2A^{-}\rho_\mathcal{S}(t)A^{+}-\{A^{+}A^{-},\rho_\mathcal{S}(t)\})\\
&+ \frac{\gamma(1+\xi)}{4}(2A^{+}\rho_\mathcal{S}(t)A^{-}-\{A^{-}A^{+},\rho_\mathcal{S}(t)\}),
\end{aligned}
\label{Eq:equationME}
\end{equation}
where $\xi =\tanh(-\beta \omega)$ is the environment correlation function and $\gamma=g^2\tau$ describes the decay rate. Moreover,  it is important to emphasis that the when the temperature of environment state is low ($\xi \rightarrow -1$), we can eventually obtain the master equation in vacuum environment,
\begin{equation}
\frac{d}{dt}\rho_\mathcal{S}(t) = -i[H_{\mathcal{S}},\rho_\mathcal{S}(t)]+\frac{\gamma}{2}(2A^{-}\rho_\mathcal{S}(t)A^{+}-\{A^{+}A^{-},\rho_\mathcal{S}(t)\}).
\label{Eq:equME}
\end{equation}

To corroborate the equivalence between the CM and quantum master equation descriptions for different temperatures, we show the stroboscopic and continuous-time evolution of $\langle\sigma_m^x\rangle$ for the subsystems in Figs. \ref{FIG:MEvsCM}(a)-(c). Note that the time line for CM has been rescaled to $t=n\tau$. One can see that the results calculated through both descriptions agree well with each other in the continuous-time limit $\tau\rightarrow 0$ regardless of the temperature of the environment. Moreover, we observe again that the synchronization of the subsystems is established when the temperature of the environment is low. In the case of the vacuum environment, the synchronized oscillations of the local observable of each subsystem are always persisted.

\subsection{Liouvillian spectrum}
So far we have verified the equivalence between the CM and master equation descriptions for the Markovian dynamics of open systems. In this section we will concentrate on the case of vacuum environment since the oscillations in the dynamics survive in the long-time limit. Actually, the Lindblad master equation Eq. (\ref{Eq:equME}) always hints a linear CPTP map, which can be described as $
\frac{d}{dt}\rho_{\mathcal{S}}(t)=\mathcal{L}[\rho_{\mathcal{S}}(t)]$ where $\mathcal{L}[\cdot]$ is the Liouvillian superoperator acting on the density matrix of the system. The Liouvillian $\mathcal{L}$  is the generator of dynamics semigroup $e^{\mathcal{L}t}$ ($t\geq0$). This reminds us to figure out the origin of the synchronization in our model via the symmetry of the Liouvilllian.

In Fock-Liouville space, the Liouvillian can be recast in a non-Hermitian matrix $\bar{\bar{\mathcal{L}}}$ as follows,
\begin{equation}
\begin{aligned}
\bar{\bar{\mathcal{L}}}=&-i\left[\left(H\otimes\mathbb{I}\right)-\left(\mathbb{I}\otimes H^{\text{T}}\right)\right]\\
&+\frac{\gamma}{2}\left(2L\otimes L^{*}-L^{\dagger}L\otimes\mathbb{I}-\mathbb{I}\otimes L^{\text{T}}L^{*}\right),
\end{aligned}
\label{Eq:LSpectrum}
\end{equation}
where $\mathbb{I}$ denotes the identity operator and $L$ is the corresponding jump operator in master equation (the superscript `T' denotes the transpose of matrix). The eigenvalue decomposition on the Liouvillian matrix reads \cite{KesslerPRA2012,MingantiPRA2018},
\begin{equation}
\bar{\bar{\mathcal{L}}}|\rho_{j}\rangle\rangle=\lambda_{j}|\rho_{j}\rangle\rangle,
\label{Eq:LSpectrum}
\end{equation}
where $|\rho_j\rangle\rangle$ are the eigenvectors, $\lambda_j$ are the associated complex eigenvalues. The real parts of the eigenvalues $\lambda_j$ are always negative-semidefinite and can be sorted in the descending order as $0\ge\text{Re}[\lambda_{0}]>\text{Re}[\lambda_{1}]>\text{Re}[\lambda_{2}]>\cdots > \text{Re}[\lambda_{n}]$. An arbitrary initial state can always be represented in a superposition of $\rho_j$ as $\rho_{\mathcal{S}}(0)=\sum_{j}{c_j\rho_j}$ where $c_j$ is the probability amplitude. Therefore, at any time $t>0$, the system evolves to the following state
\begin{equation}
\rho_{\mathcal{S}}(t)\propto\frac{\rho_{0}}{\text{tr}(\rho_{0})}+\sum_{j\neq0}c_{j}e^{\lambda_{j}t}\rho_{j}.
\label{Eq:dynaL}
\end{equation}

It is obvious that the eigenstate $\rho_0$, which is associated with the zero eigenvalues, i.e. $\text{Re}[\lambda_{0}]=\text{Im}[\lambda_{0}]=0$, will remain unchanged and is the asymptotic steady state. On the other hand, the components in the summation will vanish in the long-time limit ($t\rightarrow\infty$) because of the negative real part of $\lambda_j$. However, it is remarkable that the pure imaginary eigenvalues, $\text{Re}[\lambda_j]=0$ and $\text{Im}[\lambda_j]\ne0$, may protect the corresponding eigenstates from being decay and lead to a persistent oscillation with time. Therefore the emergence of the oscillations of the local observables stems from the appearance of the pure imaginary eigenvalues of the underlying Liouvillian superoperator.

In Figs. \ref{FIG:MEvsCM}(d)-(f), we show the Liouvillian spectra in the complex plane. One can see that the eigenvalues are always symmetrically distributed about the real axis ($\text{Im}[\lambda]=0$). We are interested in the eigenvalues with large real parts that are close to the imaginary axis. As shown in Fig. \ref{FIG:MEvsCM}(d), there exist zero eigenvalues for high temperature $\beta=1$ indicating the existence of the asymptotic steady states. Note also that zero eigenvalues have degenerated. As the temperature lowers, a pair of conjugate eigenvalues get closer to the imaginary axis as shown Fig. \ref{FIG:MEvsCM}(e). Although the dynamics of the system is yet a damped oscillation, it is a precursor of the birth of persistent oscillations in the dynamics. For $\beta=10$, we observe a pair of conjugated pure imaginary eigenvalues $\lambda=\pm2i$ which is responsible for the emergence of the persistent oscillation in the dynamics. Moreover, according to Eq. (\ref{Eq:dynaL}), we are able to deduce the frequency of the oscillation as $f=|\pm2i|/2\pi\approx0.32$ which is consistent with the FFT analysis in Fig. \ref{FIG:SyncVSCoh}(c).

\subsection{The structure of the steady-state density matrix}
As mentioned in Eq. (\ref{Eq:dynaL}), the long-time density matrix of the system is constructed of the eigenstates of Liouvillian associated to the eigenvalues with vanishing real parts. In particular, the long-time oscillating behavior of the system requires the basis matrix $M_j$ of the density matrix satisfying
\begin{equation}
[M_{j},H_{\mathcal{S}}]=m_{j}M_{j},\quad [A^{-},M_{j}]=[A^{+},M_{j}]=0,
\label{Eq:oscillatingC}
\end{equation}
where $m_j$ is real \cite{Albert2014PRA,JakschNC2019}.

In our CM, the equivalent master equation (\ref{Eq:equME}) always has a fixed solution with all spins pointing down along the $z$-direction, i.e.
$|\psi_{\downarrow}\rangle=|\downarrow\downarrow\downarrow\rangle$, corresponding the eigenenergy $\varepsilon_{\downarrow}$. Meanwhile, the Hamiltonian of the system has at least one pair of degenerate eigenstates $|\psi_{m}\rangle$ and $|\psi_{n}\rangle$ with $\varepsilon_{m}=\varepsilon_{n}$. These two degenerate eigenstates together with the state $|\psi_{\downarrow}\rangle$
can construct nine eigenoperators: $M_{1} = |\psi_{m}\rangle\langle\psi_{m}|$, $M_{2} = |\psi_{n}\rangle\langle\psi_{n}|$, $M_{3} = |\psi_{\downarrow}\rangle\langle\psi_{\downarrow}|$, $M_{4} = |\psi_{m}\rangle\langle\psi_{n}|$, $M_{5} = |\psi_{n}\rangle\langle\psi_{m}|$,
$M_{6} = |\psi_{\downarrow}\rangle\langle\psi_{m}|$, $M_{7} = |\psi_{\downarrow}\rangle\langle\psi_{n}|$, $M_{8} = |\psi_{n}\rangle\langle\psi_{\downarrow}|$ and $M_{9} = |\psi_{n}\rangle\langle\psi_{\downarrow}|$ which are fullfilled the requirements in Eq. (\ref{Eq:oscillatingC}). As a consequence, we can obtain the equation of the dynamics of those eigenoperators as
\begin{equation}
\frac{d}{dt}M_{j}=-im_{j}M_{j}=-i(\mathcal{\varepsilon}_{l}-\varepsilon_{k})|\psi_{l}\rangle\langle\psi_{k}|, \quad l,k=\downarrow,m,n. 	
\end{equation}
The matrices $M_{1},\cdots, M_{5}$ are the dark states in the dynamical process, matrices $M_{6},\cdots, M_{9}$ denote the mixed coherence oscillation. Choosing the same parameter as in Fig.\ref{FIG:MEvsCM}, we can obtain a pair of degenerate eigenstates with $p_{n}=p_{m}=3$ and $p_{\downarrow}=1$. In the long-time limit, the system eventually evolves in the eigenstate subspace consisting of $|\psi_{n}\rangle$, $|\psi_{m}\rangle$, $|\psi_{\downarrow}\rangle$, and oscillates in following ways: $|\psi_{n}\rangle\leftrightarrow|\psi_{\downarrow}\rangle$ and $|\psi_{m}\rangle\leftrightarrow|\psi_{\downarrow}\rangle$ with the frequencies $f_{n\leftrightarrow\downarrow}=f_{m\leftrightarrow\downarrow}=|p_{n,m}-p_{\downarrow}|/2\pi\approx 0.32$ in consistent with the previous results.

Actually, the conditions in Eq. (\ref{Eq:oscillatingC}) suggests that the emergence of persistent oscillations is supported by the so-called strong dynamical symmetry of the system. To be specific, in Eq. (\ref{Eq:oscillatingC}) the former commutator defines a dynamical symmetry of the autonomous evolution of the system while the latter commutator ensures that such dynamical symmetry is survived in presence of dissipation \cite{JakschNC2019,GuarnieriPRA2022}. Recently, the dynamical symmetry has been shown to be powerful in analyzing the properties of the limit cycles in the dynamics of the open quantum system, for example, the anti-synchronization in a three spin-1/2 system (with one of the spins acting as a bath) \cite{Buca2022sci}, and the robustness of the persistent oscillations in a periodically arranged four-site spin chain in a collision model with random time for the system's autonomous evolution \cite{GuarnieriPRA2022}.

In particular, as shown in Ref. \cite{GuarnieriPRA2022} the CPTP map in Eq. (\ref{Eq:CMofStateMap}) can be described by the time-evolution and Kraus operators, without taking the continuous time limit for the derivation of the Lindblian master equation, the oscillation frequency is unrelated to any time-scale of the system-environment interaction and is just of a matter of the spectrum of the CPTP map. As will be seen in the ultra-strong non-Markovian case, the failure of composition of CPTP maps in describing the time-evolution breaks the persistence oscillation in our CM in the ultra-strong non-Markovian case.

\subsection{Entropy and correlations}
\label{EntropyProduction}
In this section, we discuss the dynamics of the system from the thermodynamic viewpoint.
In the general description of the dynamics of open quantum system, the non-unitary time evolution of the state of system of interests is given by the following dynamical map, given the initial state of system as $\rho^{\text{ini}}$,
\begin{equation}
\rho_{\mathcal{S}}^{\text{ini}}\mapsto\rho_{\mathcal{S}}(t)=\text{tr}_{E}[U(t)(\rho_{\mathcal{S}}^{\text{ini}}\otimes\eta_{\text{th}})U^{\dagger}(t)],
\label{Eq:HSunitary}
\end{equation}
where $U(t)$ is the unitary time-evolution of the system and environment and $\text{tr}_E$ is the partial trace taken over the degree of freedom of the environment. In Eq. (\ref{Eq:HSunitary}), it is assumed that the system is uncorrelated with the environment at the initial time and the environment is in the thermal state characterized by a temperature. Note that a constant temperature makes sense only in the thermodynamic limit, i.e. when the number of degrees of freedom approaches infinity. In CM, such the thermal environment is represented by a set of qubits that collide with systems at each step. Thus the state of the system is stroboscopically transformed as $\rho_{\mathcal{S}}^{n-1}\rightarrow\rho_{\mathcal{S}}^{n}$ at each step which can be implemented by only involving the $n$-th environment qubit. The specific form of the time-evolution in the CM simplifies the computation of the dynamics of open systems and makes the formulation of thermodynamic in CM tractable.

It is allowed to formulate the thermodynamic in each step of the CM. For example, the entropy production $\Sigma^n$ in the $n$-th step is given by $\Sigma^n=\Delta S^n+\Phi^n$ where $\Delta S^n= S(\rho_{\mathcal{S}}^{n})-S(\rho_{\mathcal{S}}^{n-1})$ ($\rho^{\text{0}}_{S}$ is the initial state) is the entropy increment of system after time evolution with $S(\rho) = -\text{tr}(\rho\ln\rho)$ being von Neumann entropyt,  and $\Phi^{n} = \beta\Delta Q^{n}_{E}=\beta\text{tr}\left[H(\rho^{n}_{E}-\eta_{\text{th}}^{n})\right]$ is entropy flux after the $n$-th system-environment interaction. But one should pay special attention that $\Sigma^n$, $\Delta S^n$ and $\Phi^n$ are defined as the corresponding thermodynamic quantities in a single step of the CM and only the environmental qubits that couple to the system in this step can be used to calculate these thermodynamic quantities. Recall that the thermal environment in the CM consists of all the environmental qubits, a thermodynamic quantity (entropy production, work, heat, etc.) up to a time $t=n\tau$ ,with $\tau$ being the time interval between two successive collisions, should be the cumulative sum of the same quantity per step up to the $n$-th step \cite{Landi2021,strasberg2021prxquantum,purkayastha2022quantum}. Therefore the entropy production after $n$ steps of the CM reads
\begin{equation}
\Sigma(n)=\Delta S(n) + \Phi(n),
\label{Eq_entropyproduction}
\end{equation}
with the cumulated thermodynamic quantity defined by $A(n)=\sum_{j=1}^n{A^j}$. We noticed that when the persistent oscillation is absent in the long time limit the entropy production for the $n$-th step can be interpreted as the entropy production rate by simply reexpressing $\Sigma^n$ as $\Sigma^n=\Sigma(n)-\Sigma(n-1)$. It is shown that the entropy production is equal to a relative entropy and thus by construction is positive \cite{Landi2021,strasberg2021prxquantum,purkayastha2022quantum}. Nevertheless the entropy production rate need not to be positive and the Landauer's principle by means of the entropy production rate has been discussed in Refs. \cite{lorenzo2015prl,Manzhongxiao2019}. Since we are interested in the case that the persistent oscillation is present in long time limit, the physical meaning of $\Sigma^n$ is not well understood, we will focus on the cumulative quantities in the following.

\begin{figure}[!htpb]
  \includegraphics[width=1.03\linewidth]{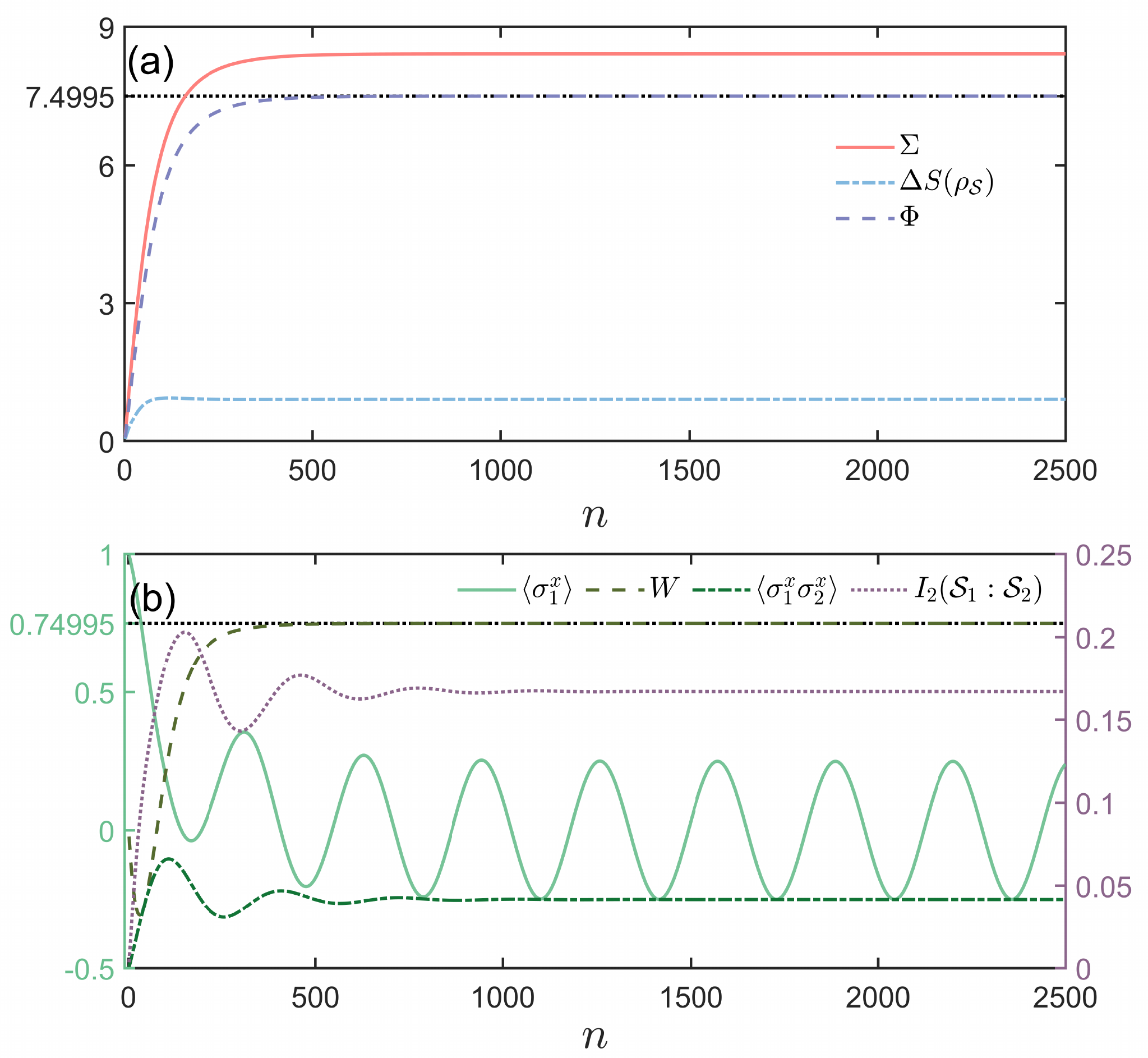}
  \caption{\label{FIG:entropyflow} (a). The $n$-dependence of the entropy production $\Sigma$, entropy increment of the system $\Delta S(\rho_S)$ and entropy flux $\Phi$, all results are obtained through Eq.(\ref{Eq_entropyproduction}). (b). The $n$-dependence of expectation $\langle\sigma^{x}_{1}\rangle$ (left $y$ axis), the work $W$ (left $y$ axis), the two-point spin-spin correlation function $\langle\sigma^{x}_{1}\sigma^{x}_{2}\rangle$ (left $y$ axis) and the bipartite mutual information of sublattices $I_{2}(\mathcal{S}_{1}:\mathcal{S}_{2})$ (right $y$ axis).  The parameters for the two panels are chosen as $J_{x}=J_{y}=-1,J_{z}=1,g=10$, $\beta=10$ and $\tau=0.01$.}
\end{figure}

In Fig.\ref{FIG:entropyflow}(a), we show the time-evolutions of the relevant thermodynamic quantities in our CM for the case in which the long-time oscillation survives. One can see that all quantities are always positive after each collision step, especially the entropy production. The always positive entropy production shows that the system always obeys Landauer's principle, the second law of thermodynamics is not violated.

In addition, the entropy production keeps increasing monotonically and remains constant after the system gradually establishes stable oscillations (combined with the results of $\langle \sigma^{x}_{1}\rangle$ in Fig.\ref{FIG:entropyflow}(b)). This can be understood as the following. On the one hand it is the coupling to the environment that leads the entropy production rises and goes to a constant (the value is proportional to the coupling strength).  On the other hand due to the dynamical symmetries, as presented in Eq. (\ref{Eq:oscillatingC}), there is a subspace of the system that is effectively decoupled from the environment throughout the time-evolution. The remaining subspace couples to the environment and approaches to an asymptotic steady state in the long-time limit. The contributions from both subspaces lead the observables oscillate persistently.

In our CM, the total energy of system is not conserved due to $[H_{\mathcal{S}},H_I]\neq0$. It means that the work $W$ may be performed during the whole evolution process \cite{Massimiliano2010,Landi2021}. According to the first law of thermodynamics, the work up to the $n$-th step is defined by the mismatch of the internal energy of the system $\Delta U$ and the heat $\Delta Q_E$ dissipated to the environment as $W (n)= \Delta U(n) + \Delta Q_E(n)$. Again the internal energy up to the $n$-th step $\Delta U(n)=\sum_{j=1}^{n}{\Delta U^j}$ is the cumulation of the change in internal energy at each step $\Delta U^j = \text{tr}\left[H_{\mathcal{S}}(\rho^{j}_{\mathcal{S}}-\rho^{j-1}_{\mathcal{S}})\right]$.

Here we show the results of the work in Fig.\ref{FIG:entropyflow}(b). The plus or minus signs of $W$ mean the energy being either poured into or extracted from the system. In Fig. \ref{FIG:entropyflow}(b), we compare the time-evolution value of $W$ and the expectation value $\langle\sigma^x_1\rangle$. One can see that at the beginning of the evolution, the value of the work is taken as negative. This behavior corresponds to the rapid dissipation of the system energy into the environment which is in accordance with the rapid decay of $\langle\sigma^{x}_{1}\rangle$. Then after a short evolutionary process up to the time $t=n\tau\approx 6 $, the work approaches to a constant rapidly. Comparing with the numerics of the steady-state value of the entropy flux in Fig. \ref{FIG:entropyflow}(a), we find that the work is equal to the heat $W = \sum_{j}\Delta Q_{E}^{j}=\Phi/\beta$, namely, the energy poured into the system happens to be completely dissipated into the environment. We note that the initial state of system is chosen as the $120^{\circ}$-state mentioned in Sec. \ref{model}. In this case, after the system undergoes long-term evolution, the internal energy of the system is completely consistent with that of the system at the initial time.


Although a subspace of the system has decoupled from its environment, the system does not enter into a stable oscillation immediately ($\langle\sigma^{x}_{1}\rangle$ has not yet reached a stable oscillation when the work reaches constant). This can be attributed to the fact that the establishment of long-time oscillation is not only induced by the dissipation but also a result of self-adjusting of the correlations among the subsystems. The evidence can be found in the time-evolutions of the correlator $\langle\sigma^{x}_{1}\sigma^{x}_{2}\rangle$ and the bipartite mutual information $I_{2}(\mathcal{S}_{1}:\mathcal{S}_{2})=S(\rho_{\mathcal{S}_{1}})+S(\rho_{\mathcal{S}_{2}})-S(\rho_{\mathcal{S}_{1}\mathcal{S}_{2}})$. As shown in Fig. \ref{FIG:entropyflow}(b), both $\langle\sigma^{x}_{1}\sigma^{x}_{2}\rangle$ and $I_{2}(\mathcal{S}_{1}:\mathcal{S}_{2})$ approach to the steady-state values simultaneously with the establishment of the stable oscillations at $n\approx 1200$.

\subsection{Imperfect interaction}
\label{sec:Imperfect}
So far, we have considered that the system-environment interaction is in the collective fashion, namely, the three system spins interact with the environment particle at the same time. However, from the experimental point of view, it is difficult to realize the simultaneous interactions between the system and the environment spins. A more realistic scenario is that the system spins interact with the environment spin sequentially in a fixed order, e.g. $\mathcal{S}_1$-$E_n\rightarrow\mathcal{S}_2$-$E_n\rightarrow\mathcal{S}_3$-$E_n$. The system-environment collision in sequential way is thus ruled by the unitary operator $U_{I,123}^{\text{seq}}=U_{\mathcal{S}_3E_n}U_{\mathcal{S}_2E_n}U_{\mathcal{S}_1E_n}$ with $U_{\mathcal{S}_mE_n}=\exp{\left[-ig\tau(\sigma_{\mathcal{S}_m}^+\sigma_{E_n}^-+\text{h.c.})\right]}$, ($m=1,2,3$). Following the expansion method mentioned in the Sec.\ref{continuoustimelimit}, we can obtain the master equation for the sequential collision case. We show the details of derivation in \hyperref[appendix]{APPENDIX}.

\begin{figure}[!htpb]
  \includegraphics[width=1\linewidth]{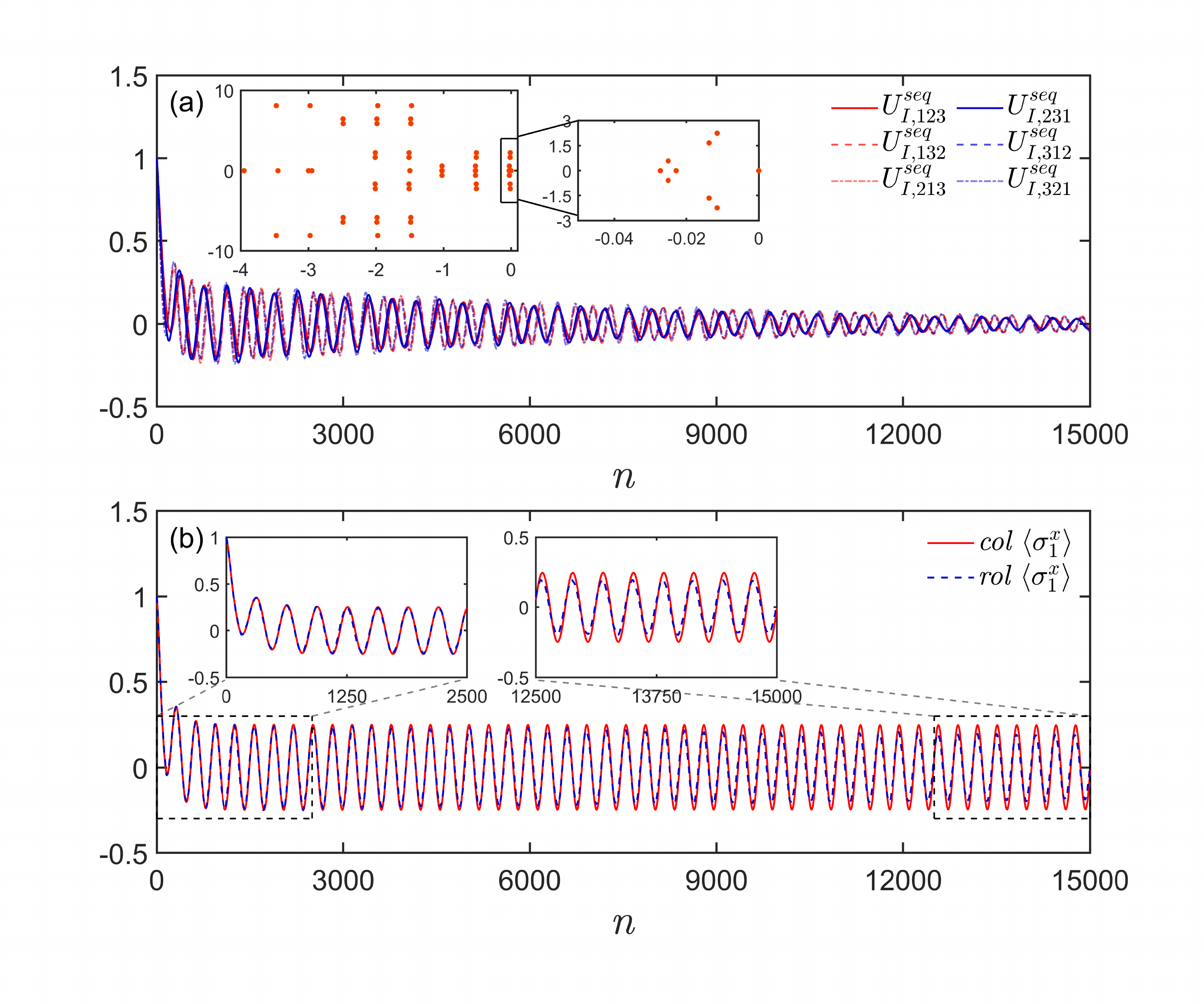}
  \caption{\label{FIG:Imperfect} (a) The stroboscopic evolution of $\langle\sigma^{x}_{1}\rangle$ in collision models with different sequences in the system-environment collision. The zoom-in shows the Liouvillian spectrum of the $U_{I,123}^{\text{seq}}$ case. (b) The stroboscopic evolution expectation values $\langle\sigma^{x}_{1}\rangle$ in the collision models with collective system-environment collision (red line) and individual subsystem-environment collisions in random order (blue line). The parameters are chosen as $J_{x}=J_{y}=-1,J_{z}=1,g=10,\tau=0.01$ and $\beta=10$.}
\end{figure}

In Fig. \ref{FIG:Imperfect}(a), we show the stroboscopic evolution of $\langle\sigma^x_1\rangle$ in the CM with system-environment interactions act sequentially in various orders (labeled by the subscripts of the time-evolution operators $U^{\text{seq}}$). In contrast to the case of collectively interactions, the oscillations of $\langle\sigma^x_1\rangle$ are observed in the initial stage of the time-evolution for all $U^{\text{seq}}$s, while as time pasts the amplitudes of the oscillations are suppressed and the magnetization $\langle\sigma^x_1\rangle$ asymptotically decay to zero. We show the Liouvillian spectrum for the case that the time-evolution operator is $U_{I,123}^{\text{seq}}$ in the inset of Fig. \ref{FIG:Imperfect}(a). One finds the largest real part of the eigenvalue of the Liouvilian to be $\text{Re}[\lambda_1]=0.0115$ and the magnetization $\langle\sigma^x_1\rangle$ asymptotically decays to zero within the time scale $\sim 1/\text{Re}[\lambda_1]$.

However, the subsystem spins interact with environment spin sequentially in a fixed order during the entire evolution is still demanding for experimental realization. We now investigate the effects of the imperfections of the interaction order on the dynamics of the system. We simulate the imperfections by randomly permuting the system spins in the queue at each step. As shown in Fig. \ref{FIG:Imperfect}(b), it is interesting that the random orders of interaction between system and environment spins significantly prolongs the oscillations and even recovers the behavior of $\langle\sigma^x_1\rangle$ appearing in the case that the system spins collectively interact with environment spin.

To address this point, we recall the sequential collision master equation shown in \hyperref[appendix]{APPENDIX}. Comparing with Eq. (\ref{Eq:equME}), one finds some additional terms appears in Eq. (\ref{eq:seqmaster}) which are determined by the collision sequence. These extra terms will be mixed destructively due to the rapid permutation of the subsystems in the system-environment interaction during the whole time-evolution. For instance,
the term $\sigma^{+}_{1}\sigma^{-}_{3}\rho_{\mathcal{S}}(t)$ appearing in the  $\mathcal{S}_1$-$E_n\rightarrow\mathcal{S}_2$-$E_n\rightarrow\mathcal{S}_3$-$E_n$ sequence cancels the term $-\sigma^{+}_{1}\sigma^{-}_{3}\rho_{\mathcal{S}}(t)$ appearing in the $\mathcal{S}_3$-$E_n\rightarrow\mathcal{S}_2$-$E_n\rightarrow\mathcal{S}_1$-$E_n$ sequence. Therefore, in the short-term evolution, the time-evolution of $\langle\sigma^x_1\rangle$ in random collision strategy agrees well with that in the collective collision strategy. However, after sufficient long time, the difference between the two cases is accumulated so that the amplitude of the oscillation differs.

\section{non-Markovian case}
\label{non-Markovian case}
We turn to the non-Markovian case in this section by switching on interaction between neighboring environmental spins.
As we mentioned in Sec.\ref{model}, in the CM framework, the non-Markovian dynamic can be implemented by introducing the inner collision between the environment spins $U_E$. The idea is the following: At the collision step $n$, the system $\mathcal{S}$ collides with the environment spin $E_{n}$, and the information from the system partially flows into the environment. Then the environment inner collision takes place, the environment spin $E_{n}$ collides with the fresh environment spin $E_{n+1}$. In this way, the spin $E_{n+1}$ also partially carries the system information. In the next collision step $n+1$, when a collision happens between the system $\mathcal{S}$ and the environment $E_{n+1}$, the information that flowed into environment has the possibility to flow back to the system.

Although the $E_n$-$E_{n+1}$ collision enables the information backflow, a measure still need to quantify the degree of the non-Markovianity of the dynamics of the system. To this aim, we employ the well-known BLP measure \cite{Breuer2009PRL} and generalize it to the discrete time evolution.

\begin{figure}[!htpb]
  \includegraphics[width=1\linewidth]{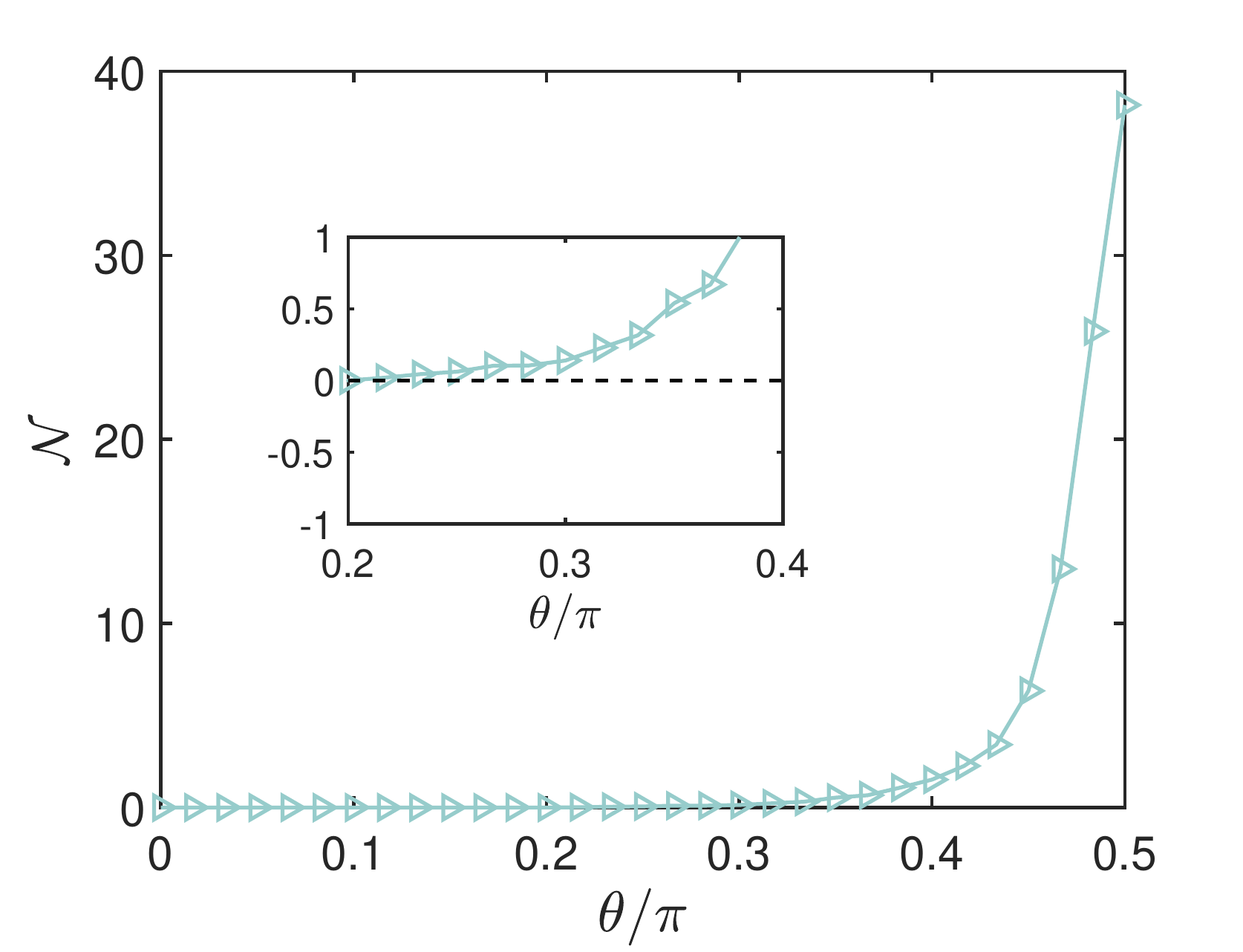}
  \caption{\label{FIG:D} The degree of non-Markovianity as a function of $\theta/\pi$. The number of collisions is $n=500$. Other parameters are chosen as $J_{x}=J_{y}=-1$, $J_{z}=1$, $\beta=10$, $g=5$ and $\tau=0.08$.}
\end{figure}

The idea of the BLP measure is to quantify the non-Markovianity through the trace distance change of the system state as follows,
\begin{equation}
\mathcal{N}=\max_{\rho_{1},\rho_{2}}\sum_{\Delta\mathcal{D}(n)>0} \Delta\mathcal{D}(n),
\label{Eq:tracedistanceNM}
\end{equation}
with $\Delta\mathcal{D}(n)=\mathcal{D}[\rho_{1}(n+1),\rho_{2}(n+1)]-\mathcal{D}[\rho_{1}(n),\rho_{2}(n)]$
and $\mathcal{D}(\rho_{1},\rho_{2})=\text{tr}(\sqrt{(\rho_{1}-\rho_{2})^\dagger(\rho_{1}-\rho_{2}})/2$ is the trace distance between the two density matrices $\rho_{1}$ and $\rho_{2}$. The trace distance quantifies the degree of distinguishability between the two states with $0\leq\mathcal{D}(\rho_{1},\rho_{2})\leq1$ and $\mathcal{D}(\rho_{1},\rho_{2})=1$ corresponds to the case where the two states are orthogonal.
In the Markovian process, because the information of the system is one-way flows into the environment, all distinct initial states will eventually approach to a unique steady state manifested by $\Delta\mathcal{D}(n)<0$ for all $n$. However, if it exists $\Delta\mathcal{D}(n)>0$ for some $n$, namely the distinguishability of the two initial states increases, it is signal that the information have flowed back to the system at some point thus the dynamics is non-Markovian. The non-Markovianity is obtained by the sum of all the increasing of the trace distance during the time-evolution.

In principle, the maximization in Eq. (\ref{Eq:tracedistanceNM}) should run over all the possible initial states. Here we performed $K=200$ simulations with $\rho_{1}$ being a random state and $\rho_{2} = \rho_{1}^{\perp}$ being the orthogonal state of $\rho_1$. The non-Markovianity $\mathcal{N}$ is thus given by the optimal pair of $\rho_1$ and $\rho_2$ within the all performed simulations. The non-Markovianity $\mathcal{N}$ as a function of the strength of $E_n$-$E_{n+1}$ collision $\theta$ is shown in Fig.\ref{FIG:D}. Note that according to the equation $\theta=2g_{E}\tau$ in Sec.\ref{model}, the coupling strength between environmental spins is proportional to $\theta$. One can find that the non-Markovianity of the dynamics is not induced as soon as the $E_n$-$E_{n+1}$ collision is switched on. When the $E_n$-$E_{n+1}$ collision is weak, e.g. $\theta/\pi<0.2$, after $n=500$ collisions, the non-Markovianity $\mathcal{N}$ is always zero. As we continue to increase the strength of $E_n$-$E_{n+1}$ collision, the dynamics becomes non-Markovian and $\mathcal{N}$ increases monotonic with $\theta$ increasing.

\begin{figure*}[!htpb]
  \includegraphics[width=1\linewidth]{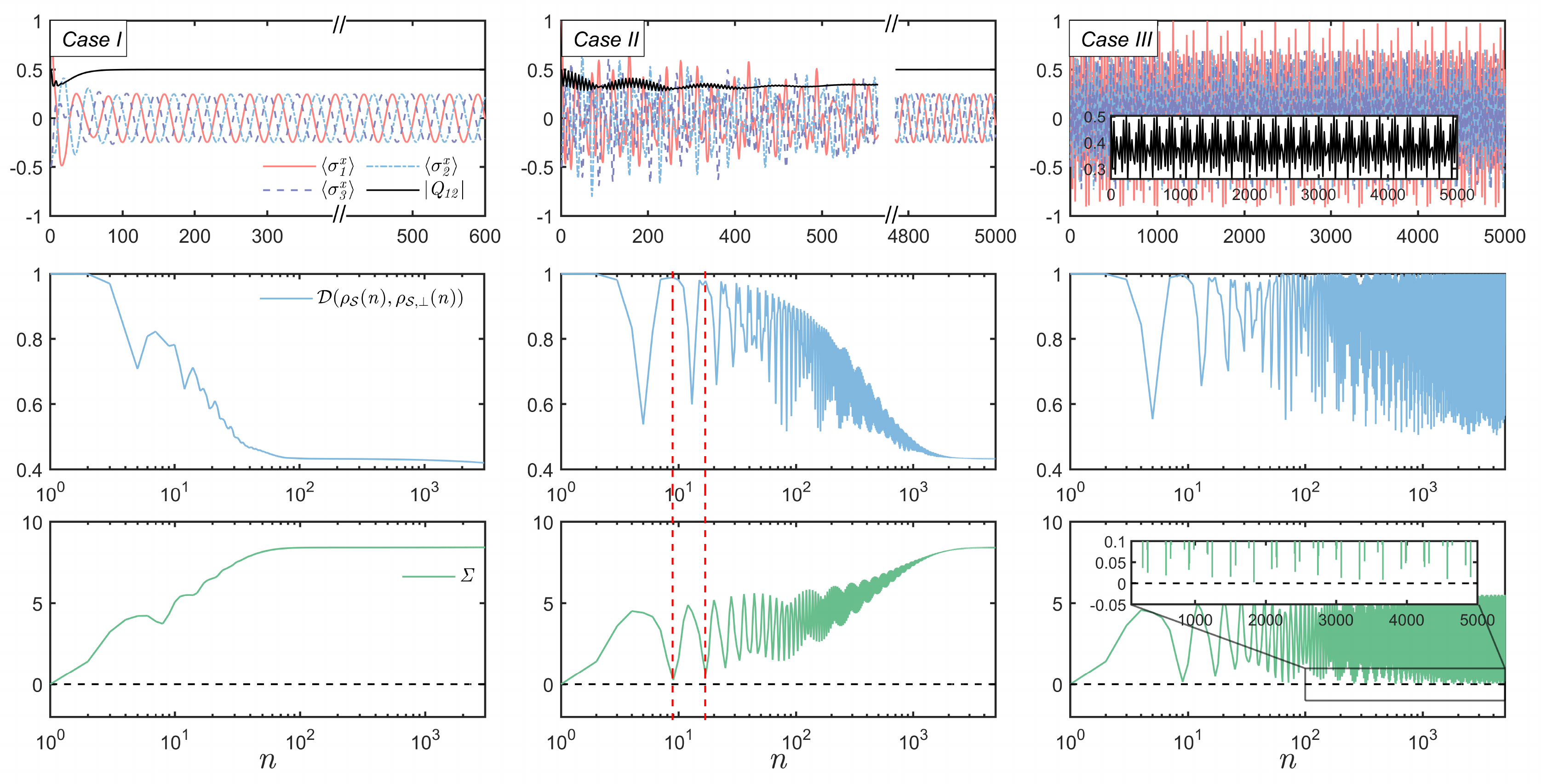}
  \caption{\label{FIG:NMCase} The time-evolution of expectation values of $\langle\sigma^{x}_{m}\rangle(m=1,2,3)$ and the modulus of the synchronization measure $Q_{12}$ (top panels), the trace distance $\mathcal{D}[\rho_{1}(n),\rho_{2}(n)]$ (middle panels) and entropy production $\Sigma$ (bottom panels) for {\it Case I}: weak non-Markovian case with $\theta=\pi/3$; {\it Case II}: strong non-Markovian case with $\theta=\pi/2.1$ and {\it Case III}: ultrastrong non-Markovian case with $\theta=\pi/2$. The initial state of the system are chosen as $\rho_1$ to be the $120^\circ$-state and $\rho_{2}$ to be the orthogonal state of $\rho_1$, i.e. $\mathcal{D}(\rho^{\text{ini}}_{\mathcal{S}},\rho^{\text{ini}}_{\mathcal{S},\perp})=1$.  Other parameters are chosen as $J_{x}=J_{y}=-1$, $J_{z}=1$, $\beta=10$, $g=5$, and $\tau=0.08$.}
\end{figure*}

For the non-Markovian region ($\theta/\pi=0.2$), we discuss through the following three cases:

\noindent{\it Case I}. The weak non-Markovian case with $\theta=\pi/3$;

\noindent{\it Case II}. The strong non-Markovian case with $\theta=\pi/2.1$;

\noindent{\it Case III}. The ultrastrong non-Markovian case with $\theta=\pi/2$.

Meanwhile, the unitary evolution operator within the environment blocks $U_{E_{n}E_{n+1}}$ is the SWAP gate when $\theta=\pi/2$. The state of the system is initialized to the $120^\circ$-state. As for initial states of the system in the calculation of trace distance $\mathcal{D}(\rho_{1},\rho_{2})$,  the state $\rho_{1}$ is the $120^\circ$-state and the other one is the orthogonal state of $\rho_{2}=\rho_1^{\perp}$.

Since the correlation between environment blocks has been established in the non-Markovian case, we have to redefine the entropy production before proceeding to a specific discussion. What is noteworthy in the non-Markovian case is that in each collision cycle, there involves two environment blocks. To be more precisely, the detail expression of interval entropy flux in each collision cycle has to contain the total energy of those two environment parts \cite{PezzuttoNJP2016}, that is
\begin{equation}
\Phi^{n} = \beta \Delta Q_E^{n} = \beta \text{tr}\left[\left(H_n \otimes H_{n+1}\right) \left(\rho_{E_nE_{n+1}}^{\text{post}} - \rho_{E_nE_{n+1}}^{\text{pre}}\right)\right].
\end{equation}
For the sake of understanding, let us recall the entail collision evolution introduced before
\begin{equation}
\rho_{\mathcal{S}E_nE_{n+1}}^{\text{pre}}\equiv\rho_{\mathcal{S}E_n} \otimes \eta_{\text{n}}^{n+1},\,\rho_{\mathcal{S}E_nE_{n+1}}^{\text{post}} \equiv U_EU_IU_{\mathcal{S}}  \rho_{\mathcal{S}E_nE_{n+1}}^{\text{pre}} U_{\mathcal{S}}^\dagger U_I^\dagger U_E^\dagger.
\end{equation}
Here once again, to avoid misunderstanding, we emphasize $H$ is the free Hamiltonian of the thermal environment particle.

In Fig. \ref{FIG:NMCase}, we show the time-evolution of $\langle\sigma^x_m\rangle$ and the modulus of the synchronization measure $Q_{12}$. For the weak non-Markovian case, one can find that the dynamics of the system do not suffer from non-Markovianity and oscillations can be rapidly built up in the evolution. As shown in the left-middle panel of Fig. \ref{FIG:NMCase}, although the dynamics is non-Markovian, the information backflow ($\Delta\mathcal{D} > 0 $) only occurs in the early stage of the evolution and then the trace distance decreases monotonically. As for the entropy production, in {\it Case I}, the entropy production is always a positive value, which is far from zero value and rapidly reaches a constant. The phenomena are similar with the Markovian case, a subspace of the system will rapidly decouple will rapidly decouple from the environment during evolution and the weak non-Markovianity does not have a significant effect on the system dynamics.

In {\it Cases II}, by contrast, both the expectation values of $\langle\sigma_m^x\rangle$ and $|Q_{12}|$ show irregular oscillations at the beginning of the evolution, see the second column of Fig. \ref{FIG:NMCase}. In spite of the early-stage haphazard behavior, the system evolution gradually becomes regular and stabilizes around $n = 2500$ (not shown in Fig.\ref{FIG:NMCase}), and the system reaches stable oscillations eventually.  The modulus of the quantum synchronization measure $|Q_{12}|$ also reaches a constant value in the long-time evolution which hints at the existence of good synchronization. The time dependence of trace distance for {\it Case II} shows a rapid oscillation until $n>1000$ which is a typical evidence for the strong non-Markovian dynamics. In particular, at some moments (marked by red dashed lines), the trace distance instantaneously recovers the initial value.

The strong non-Markovianity significantly impacts the entropy production in the time-evolution. One can see that the entropy production does not monotonically increase and may decrease drastically close to to zero at some $n$ in the discrete time-evolution. This indicates the entropy production at the $n$-th step $\Sigma^{n}$ becomes negative. This can be understood by, for example, focusing on the steps $n=9$ and $17$ ( the red dashed lines in Fig. \ref{FIG:NMCase}) at which the trace distance (almost) recovers to unity, it is the strong information backflow that diminishes the vague of different states and decrease the entropy of the system and thus leads to the negative $\Sigma^n$ at $n=9$ and $17$. But when we look at $\Sigma(n)$ through the entire time-evolution, the boundary of violation of the second law of thermodynamics is untouchable \cite{PezzuttoNJP2016}.

In {\it Case III}, the collision between neighboring environmental spins is set to be the SWAP operation. As shown in the right column of Fig.\ref{FIG:NMCase}, one can find that the dynamics of the system always exhibit irregular oscillations even in the long-time limit $(n \sim 5000)$.
In addition, the measure $|Q_{12}|$ fails to reach an asymptotic steady-state value implying the absence of synchronization in such ultrastrong non-Markovian dynamics. The time-dependence of both the trace distance as well as the entropy production oscillate irregularly. Again the non-monotonic behavior of the entropy production is observed in the ultrastrong non-Markovian case. The negative entropy production $\Sigma^n$ during the time-evolution is due to the creation of the correlations in the system and environmental qubits  \cite{Manzhongxiao2019,PezzuttoNJP2016}.

We would finish by emphasizing again that, in CM scenario, since the thermal environment is represented by a series of qubits, the thermodynamic quantities should be the cumulative of the same quantities at each step to ensure all the changes in the involved environmental qubits are taken into account. In this sense, the second law of thermodynamics still holds in our model from a macroscopic point of view.

\section{Summary}
\label{Summary}
In summary, we have investigated the dissipative dynamics of a tripartite spin system in the framework of CM. With the introduction of successive collisions between the system and environment spins, collective dissipation has been simulated. We have found that when the environment temperature is low, the dynamics of the system exhibit a well-defined oscillation and the mutual synchronization can be established among the subsystems. We proceeded the discussion by classifying the present CM into the Markovian and the non-Markovian cases, according to allow or not the collision between neighboring environmental spins.

In the Markovian case, we have discussed the effects of the coupling strength and environment temperature on the dynamics of the systems as well as the property of synchronization among subsystems. We find that the dynamics of system may show periodical oscillations when the environmental temperature is low and the anisotropy of the interactions between the subsystems tends to destroy the stable oscillation. In addition, we have extracted the frequency of the oscillation by means of FFT.

To understand the establishment of the stable oscillations in the Markovian dynamics, we have connected the descriptions of CM and master equation by virtue of the taking the continuous time limit on the interval collisions. We then analyzed the Liouvillian spectrum of the master equation and found that there indeed exist purely imaginary eigenvalues when the environment temperature is low. In particular the inverse of the modulus of the pure imaginary eigenvalues is consistent with oscillation frequency accessed by the FFT. We further investigated the structure of the density matrix of the system and found that the system's oscillatory behavior is actually a process of leapfrogging between degenerate Hamiltonian eigenstates and other energy eigenstates within a subspace consisted of eigenoperators. Finally, we investigated the temporal behaviors of entropy production and quantum correlations which shows the oscillation of the system is induced not only by the coupling to the environment but also a result of self-adjusting of the correlations among the subsystems. We also discussed the effects of the imperfection of the collective interactions on the long-time oscillations.

For the non-Markovian case, we utilized the BLP measure of non-Markovianity in the stroboscopic time-evolution in the CM. We found in the strong (and ultrastrong) non-Markovian dynamics, the entropy production is always positive and may exhibit nomonotonic behavior in the time-evolution implying that the entropy production in some collisions steps become negative. This can be understood by the fact that the backflow of information in strong non-Markovianity dynamics compensates the energy in erasing the information.  Finally, the dynamic of the system exhibits chaos-like behavior even after sufficient long time in the ultrastrong non-Markovian case. In the future work, it will be interesting to check the chaotic behavior of the system and investigate the influence of the system environment correlations on the system dynamics behavior.

\section*{ACKNOWLEDGMENTS}

This work is supported by National Natural Science Foundation of China under Grant No. 11975064.

\section*{APPENDIX}
\label{appendix}

In this appendix, we take the collision sequence given in the previous Sec.\ref{sec:Imperfect} as an example to show the specific form of the master equation for sequential collision case. Here we recall the unitary-evolution operator $U_{I,123}^{\text{seq}}=U_{\mathcal{S}_3E_n}U_{\mathcal{S}_2E_n}U_{\mathcal{S}_1E_n}$, with $U_{\mathcal{S}_mE_n}=\exp{[-ig\tau(\sigma_{\mathcal{S}_m}^+\sigma_{E_n}^-+\text{h.c.})]}$, ($m=1,2,3$) and the environment particles are still located in the thermal state without off-diagonal matrix elements.

The dynamical map for the system also can be given by,
\begin{equation}
\rho_{\mathcal{S}}^{n}\mapsto \rho_{\mathcal{S}}^{n+1} = \text{tr}_{E_{n+1}}\left[U_{I,123}^{\text{seq}}U_{\mathcal{S}} (\rho^{n}_{\mathcal{S}} \otimes \eta^{n+1}_{\text{th}})U_{\mathcal{S}}^{\dagger}U_{I,123}^{\text{seq},\dagger}\right]=\Lambda[\rho^{n}_{\mathcal{S}}].
\label{Eq:CMofState}
\end{equation}
Based on the requirements of the continuous time limit, which hints $\tau\to0$, we still expand the unitary operators as follows,
\begin{equation}
\begin{aligned}
U_{\mathcal{S}_mE_n}&=\mathbb{I}-i\tau V_{m}-\tau^2 V_{m}^2/2+o(\tau^n),\\
&\approx\mathbb{I}-ig\tau(\sigma_{\mathcal{S}_m}^+\sigma_{E_n}^-+\text{h.c.})-g^2\tau^2 (\sigma_{\mathcal{S}_m}^+\sigma_{E_n}^-+\text{h.c.})^2/2,	
\end{aligned}
\label{Use_exp}
\end{equation}
and
\begin{equation}
U_{\mathcal{S}}=\mathbb{I}-i\tau H_{\mathcal{S}}+o(\tau^n).
\label{Uss_exp}
\end{equation}

Then we can directly obtain the mapping of the collision,
\begin{widetext}
\begin{equation}
\begin{aligned}
\rho_{\mathcal{S}}^{n+1} &=\Lambda[\rho^{n}_{\mathcal{S}}]= \text{tr}_{E_{n}}\left[U_{\mathcal{S}_3E_n}U_{\mathcal{S}_2E_n}U_{\mathcal{S}_1E_n}U_{\mathcal{S}} (\rho^{n}_{\mathcal{S}} \otimes \eta^{n+1}_{\text{th}})U_{\mathcal{S}}^{\dagger}U_{\mathcal{S}_1E_n}U_{\mathcal{S}_2E_n}U_{\mathcal{S}_3E_n}\right]\\
&=\text{tr}_{E_{n}}\left[\{\mathbb{I}-i\tau(V_1+V_2+V_3 + H_{\mathcal{S}})-\frac{\tau^{2}}{2}(V^{2}_{1} + V^{2}_{2} + V^{2}_{3}) -\tau^2(V_{3}V_{2} + V_{2}V_{1} + V_{3}V_{1})-\tau^{2}(V_{3} + V_{2} + V_{1})H_{\mathcal{S}}\}\cdot \{\rho^{n}_{\mathcal{S}} \otimes \eta^{n+1}_{\text{th}} \}\cdot \right.\ \\
&\left.\quad\quad\quad\{\mathbb{I}+i\tau(V_1+V_2+V_3 + H_{\mathcal{S}})-\frac{\tau^{2}}{2}(V^{2}_{1} + V^{2}_{2} + V^{2}_{3}) -\tau^2(V_{2}V_{3} + V_{1}V_{2} + V_{1}V_{3})-\tau^{2}H_{\mathcal{S}}(V_{3} + V_{2} + V_{1})\}\right].
\end{aligned}
\end{equation}

Recall the time scale $\tau\to0$ and $g^2\tau=\text{constant}$, we have
\begin{equation}
\begin{aligned}
\rho_{\mathcal{S}}^{n+1} =&\text{tr}_{E_{n}}\left[\rho^{n}_{\mathcal{S}} \otimes \eta^{n+1}_{\text{th}}-i\tau H_{\mathcal{S}}(\rho^{n}_{\mathcal{S}} \otimes \eta^{n+1}_{\text{th}})-\frac{\tau^{2}}{2}(V^{2}_{1} + V^{2}_{2} + V^{2}_{3})(\rho^{n}_{\mathcal{S}} \otimes \eta^{n+1}_{\text{th}}) -\tau^2(V_{3}V_{2} + V_{2}V_{1} + V_{3}V_{1})(\rho^{n}_{\mathcal{S}} \otimes \eta^{n+1}_{\text{th}}) \right.\\
&+\left. i\tau(\rho^{n}_{\mathcal{S}} \otimes \eta^{n+1}_{\text{th}}) H_{\mathcal{S}}+ \tau^{2}(V_{3} + V_{2} + V_{1})(\rho^{n}_{\mathcal{S}} \otimes \eta^{n+1}_{\text{th}})(V_1+V_2+V_3)-\frac{\tau^{2}}{2}(\rho^{n}_{\mathcal{S}} \otimes \eta^{n+1}_{\text{th}})(V^{2}_{1} + V^{2}_{2} + V^{2}_{3})\right].\\
=&\rho^{n}_{\mathcal{S}}-i\tau[H_{\mathcal{S}},\rho^{n}_{\mathcal{S}}]+\frac{g^{2}\tau^{2}}{2}\sum_{m}\langle\sigma^{+}_{E_{n}}\sigma^{+}_{E_{n}}\rangle_{\eta^{n+1}_{\text{th}}}(2\sigma^{-}_{m}\rho^{n}_{\mathcal{S}}\sigma^{-}_{m} - \{\sigma^{-}_{m}\sigma^{-}_{m},\rho^{n}_{\mathcal{S}}\})+\frac{g^{2}\tau^{2}}{2}\sum_{m}\langle\sigma^{-}_{E_{n}}\sigma^{-}_{E_{n}}\rangle_{\eta^{n+1}_{\text{th}}}(2\sigma^{+}_{m}\rho^{n}_{\mathcal{S}}\sigma^{+}_{m} - \{\sigma^{+}_{m}\sigma^{+}_{m},\rho^{n}_{\mathcal{S}}\}) \\
&+\frac{g^{2}\tau^{2}}{2}\sum_{m}\langle\sigma^{-}_{E_{n}}\sigma^{+}_{E_{n}}\rangle_{\eta^{n+1}_{\text{th}}}(2\sigma^{-}_{m}\rho^{n}_{\mathcal{S}}\sigma^{+}_{m} - \{\sigma^{+}_{m}\sigma^{-}_{m},\rho^{n}_{\mathcal{S}}\}) + \frac{g^{2}\tau^{2}}{2}\sum_{m}\langle\sigma^{+}_{E_{n}}\sigma^{-}_{E_{n}}\rangle_{\eta^{n+1}_{\text{th}}}(2\sigma^{+}_{m}\rho^{n}_{\mathcal{S}}\sigma^{-}_{m} - \{\sigma^{-}_{m}\sigma^{+}_{m},\rho^{n}_{\mathcal{S}}\})\\
&+\frac{g^{2}\tau^{2}}{2}\sum_{m\neq k}\langle\sigma^{+}_{E_{n}}\sigma^{+}_{E_{n}}\rangle_{\eta^{n+1}_{\text{th}}}(2\sigma^{-}_{m}\rho^{n}_{\mathcal{S}}\sigma^{-}_{k} - \sigma^{-}_{m}\sigma^{-}_{k}\rho^{n}_{\mathcal{S}}- \rho^{n}_{\mathcal{S}}\sigma^{-}_{k}\sigma^{-}_{m} ) + \frac{g^{2}\tau^{2}}{2}\sum_{m\neq k}\langle\sigma^{-}_{E_{n}}\sigma^{-}_{E_{n}}\rangle_{\eta^{n+1}_{\text{th}}}(2\sigma^{+}_{m}\rho^{n}_{\mathcal{S}}\sigma^{+}_{k} - \sigma^{+}_{m}\sigma^{+}_{k}\rho^{n}_{\mathcal{S}}- \rho^{n}_{\mathcal{S}}\sigma^{+}_{k}\sigma^{+}_{m} ) \\
&+\frac{g^{2}\tau^{2}}{2}\sum_{m>k}\langle\sigma^{+}_{E_{n}}\sigma^{-}_{E_{n}}\rangle_{\eta^{n+1}_{\text{th}}}(2\sigma^{+}_{m}\rho^{n}_{\mathcal{S}}\sigma^{-}_{k} - \sigma^{-}_{m}\sigma^{+}_{k}\rho^{n}_{\mathcal{S}}- \rho^{n}_{\mathcal{S}}\sigma^{-}_{k}\sigma^{+}_{m} ) +\frac{g^{2}\tau^{2}}{2}\sum_{m>k}\langle\sigma^{-}_{E_{n}}\sigma^{+}_{E_{n}}\rangle_{\eta^{n+1}_{\text{th}}}(2\sigma^{-}_{m}\rho^{n}_{\mathcal{S}}\sigma^{+}_{k} - \sigma^{+}_{m}\sigma^{-}_{k}\rho^{n}_{\mathcal{S}}- \rho^{n}_{\mathcal{S}}\sigma^{+}_{k}\sigma^{-}_{m} )\\
&+\frac{g^{2}\tau^{2}}{2}\sum_{m<k}\langle\sigma^{-}_{E_{n}}\sigma^{+}_{E_{n}}\rangle_{\eta^{n+1}_{\text{th}}}(2\sigma^{+}_{m}\rho^{n}_{\mathcal{S}}\sigma^{-}_{k} - \sigma^{-}_{m}\sigma^{+}_{k}\rho^{n}_{\mathcal{S}}- \rho^{n}_{\mathcal{S}}\sigma^{-}_{k}\sigma^{+}_{m} ) +\frac{g^{2}\tau^{2}}{2}\sum_{m<k}\langle\sigma^{+}_{E_{n}}\sigma^{-}_{E_{n}}\rangle_{\eta^{n+1}_{\text{th}}}(2\sigma^{-}_{m}\rho^{n}_{\mathcal{S}}\sigma^{+}_{k} - \sigma^{+}_{m}\sigma^{-}_{k}\rho^{n}_{\mathcal{S}}- \rho^{n}_{\mathcal{S}}\sigma^{+}_{k}\sigma^{-}_{m} ).
\end{aligned}
\label{Eq:AppMS}
\end{equation}
\end{widetext}

\begin{figure}[!htpb]
  \includegraphics[width=1\linewidth]{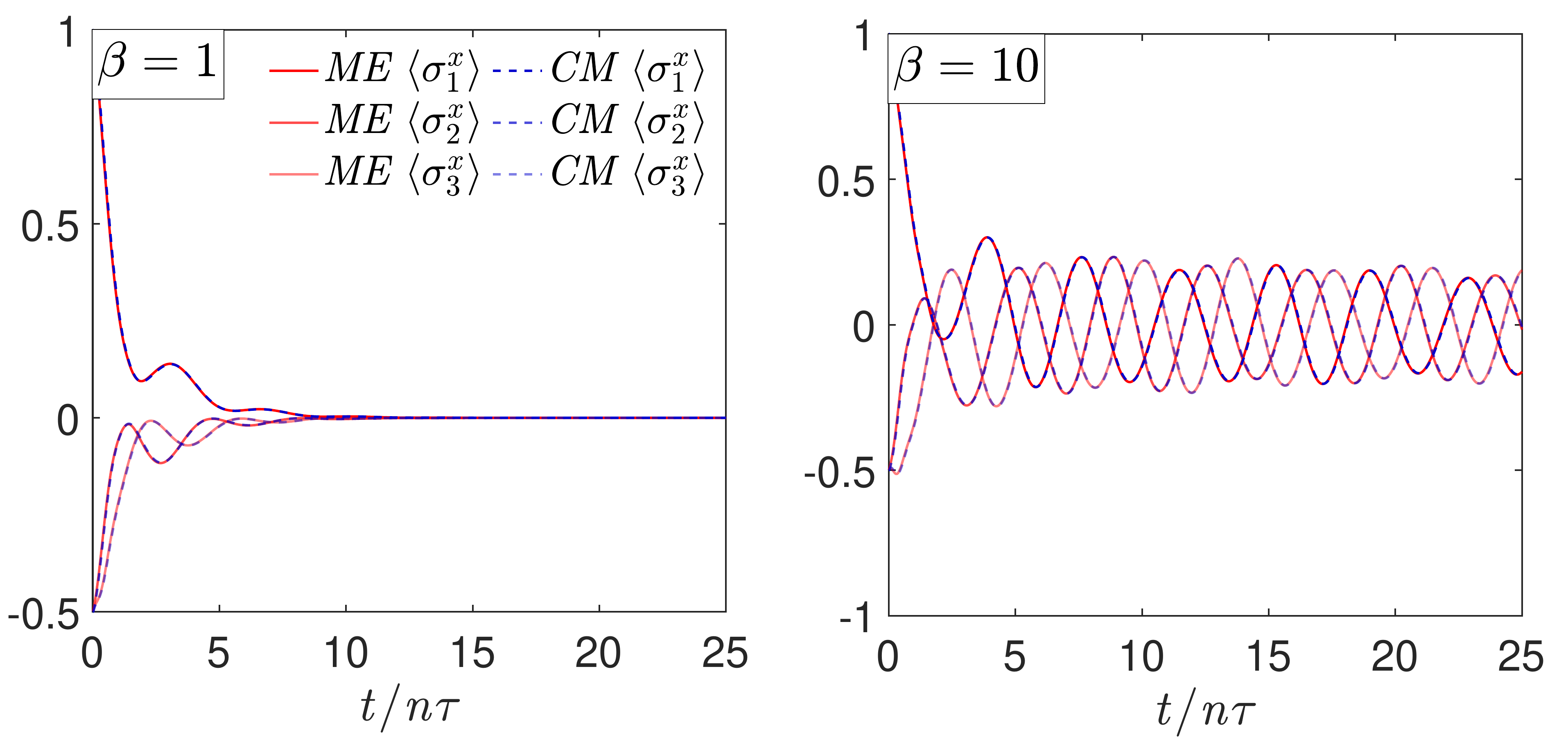}
  \caption{\label{FIG:Appendix} The time-evolution of $\langle\sigma^{x}_{m}\rangle$ $(m = 1, 2, 3)$ of the system $\mathcal{S}$ (the subsystems are distinguished by the transparency of the lines) in the description of master equation (\ref{Eq:AppMS}) and collision model for $\beta=1$ (left) and $10$ (right). Other parameters are chosen as $J_{x}=J_{y}=-1,J_{z}=1,g=10$ and $\tau=0.01$.}
\end{figure}
In the above equation, the terms with $\tau^{2}$ and $g\tau^{2}$ are all neglected, and the summations runs over $m, k=1,2,3$. We have to especially emphasis that, the relation between $m$ and $k$ in the summations is related to and satisfies the current collision sequence. Then we again emphasize the specific form of state of environment particle, the state without the off-diagonal elements will lead to $\langle\sigma^{+}_{E_{n}}\sigma^{+}_{E_{n}}\rangle_{\eta^{n+1}_{\text{th}}}=\langle\sigma^{-}_{E_{n}}\sigma^{-}_{E_{n}}\rangle_{\eta^{n+1}_{\text{th}}}=0$, and the results of $\langle\sigma^{-}_{E_{n}}\sigma^{+}_{E_{n}}\rangle_{\eta^{n+1}_{\text{th}}}=(1-\xi)/2$ and $\langle\sigma^{+}_{E_{n}}\sigma^{-}_{E_{n}}\rangle_{\eta^{n+1}_{\text{th}}}=(1+\xi)/2$ with $\xi =\tanh(-\beta \omega)$. The parameters $\beta$ and $\omega$ are environmental conditions which we have mentioned in Sec.\ref{sec:Imperfect}. We also present the comparison of above master equation and CM in Fig.\ref{FIG:Appendix}. Following the identical method in previous section, and we still consider the environment state in low temperature, e.g. $\xi\approx-1$, we can finally obtain the master equation,
\begin{eqnarray}
\frac{d}{dt}\rho_\mathcal{S}(t) &= & -i[H_{\mathcal{S}},\rho_\mathcal{S}(t)]+ \frac{\gamma}{2}\left(2 A^{-} \rho_{\mathcal{S}}(t) A^{+} - \{ \rho_{\mathcal{S}} (t), A^{+} A^{-}\}\right) \cr\cr&&+\frac{\gamma}{2}[\sigma^{+}_{2}\sigma^{-}_{3}+\sigma^{+}_{1}\sigma^{-}_{3}+\sigma^{+}_{1}\sigma^{-}_{2},\rho_\mathcal{S}(t)]\cr\cr
&&+\frac{\gamma}{2}[\rho_\mathcal{S}(t),\sigma^{+}_{3}\sigma^{-}_{2}+\sigma^{+}_{3}\sigma^{-}_{1}+\sigma^{+}_{2}\sigma^{-}_{1}],
\label{eq:seqmaster}
\end{eqnarray}
where $A^{\pm}=\sum_{m}\sigma^{\pm}_{m}$ are the collective lowering and raising operators, the corresponding Liouvillian spectrum is shown in the zoom-in of Fig.\ref{FIG:Imperfect}, which hints that the system is unable to govern the stable oscillates.

\end{document}